\pdfoutput=1
\documentclass[aps,prb,showpacs,twocolumn,eqsecnum]{revtex4-1}
\usepackage[utf8]{inputenc}
\usepackage{amsmath}
\usepackage{amssymb}
\usepackage{graphicx}
\usepackage{color}
\usepackage{bm}

\def\y{u}

\begin{document}
\title{Scaling of entanglement  in  
$2+1$-dimensional scale-invariant  field theories}
\author{Xiao Chen, Gil Young Cho, Thomas Faulkner, and Eduardo Fradkin}
\affiliation{Department of Physics and Institute for Condensed Matter Theory, University of Illinois at Urbana-Champaign, 
Illinois 61801-3080, USA}

\date{\today}
\begin{abstract}
We study the universal scaling behavior of the entanglement entropy of critical theories in $2+1$ dimensions. We specially consider two fermionic scale-invariant  models,  free massless Dirac fermions and a model of fermions with quadratic band touching, and numerically study the two-cylinder entanglement entropy of the models on the torus. We find that in both cases the entanglement entropy satisfies the area law and has the subleading term which is a scaling function of the aspect ratios of the cylindrical regions.  
We test the scaling of entanglement in both the free fermion models using three possible scaling functions for the subleading term derived from a) the quasi-one-dimensional conformal field theory, b) the bosonic quantum Lifshitz model, and c) the holographic AdS/CFT correspondence. For the later case we construct an analytic scaling function using holography, appropriate for critical theories with a gravitational dual description.
We find that the subleading term in the fermionic models is well described, for a range of aspect ratios, 
by the scaling form 
derived from the quantum Lifshitz model  as well as that derived using the AdS/CFT correspondence (in this case only for the Dirac model). For the case where the fermionic models are placed on a square torus we find the fit to the different scaling forms is in agreement to surprisingly high precision.
\end{abstract} 

\maketitle

\section{Introduction}

The von Neumann entanglement entropy (EE) has  proven to be a useful tool to diagnose and characterize strongly coupled field theories and condensed matter systems such as the topologically ordered phases and quantum critical systems.
The von Neumann EE $S_{vN}$ in a massive phase is well understood and it has been shown to satisfy an area law, $S_{vN}=\alpha \left(\frac{\ell}{\epsilon}\right)^{d-1}$, where $d$ is the dimension of space, $\ell$ is the linear size of the region $A$ being observed, and $\alpha$ is a non-universal (cutoff-dependent) constant.\cite{Bombelli-1986,Srednicki-1993,Wolf-2008} The EE of massless and in generally scale-invariant field theories in spatial dimensions $d>1$ is also expected  to obey the area law since it reflects the short-range entanglement generally present in the ground-state wave functions of these local field theories. This expectation is confirmed by the general result derived from the AdS/CFT correspondence in relativistic scale-invariant theories,\cite{Ryu2006}  in calculations in free-field theories,\cite{Casini-2009} and in many models in condensed matter physics in one and two spatial dimensions.\cite{Amico2008,Eisert-2010,Fradkin2013}

A rather special situation occurs in one-dimensional quantum critical systems and $1+1$-dimensional conformal field theories (CFT), where it was shown that the von Neumann EE of a subregion $A$ of a partition $A \cup B$ has the universal form\cite{Callan1994,Holzhey1994,Calabrese2004,Calabrese-2009b} $S_{vN}= \frac{c}{3}\log \frac{\ell}{\epsilon}$, where $c$ is the central charge of the CFT, $\epsilon$ is the short-distance cutoff, and $\ell$ is the length of a large subregion $A$ (provided $\epsilon \ll \ell \ll L$, where  $L$ is the linear size of the system).  Although formally this logarithmic scaling law is consistent with the formal $d\to 1$ limit of the area law, it represents the long-range entanglement properties of $1+1$-dimensional CFTs instead of the short-ranged entanglement reflected in the area law for $d >1$. Finite sub-leading contributions (as a function of the size of the region) to the EE (in the form of  multi-region partitions,  mutual information and entanglement negativity) are also known to be determined by the structure of the CFTs and reflect the structure of the large-scale entanglement encoded in their ground state wave functions.\cite{Caraglio-2008,Calabrese-2009,Calabrese2012,Cardy-2013}

Large-scale entanglement is expected to be found in topological phases of matter (described by topological field theories) and in scale-invariant field theories (which are at a fixed point of the renormalization group). In the cases of topological phases and topological quantum field theories in two spatial dimensions, the EE was shown to obey the scaling law $S_{vN}=\alpha \frac{\ell}{\epsilon}-\gamma_{\rm topo}$, where $\alpha$ is non-universal and $\ell$ is the linear size of the macroscopic subregion $A$, and for a topologically-trivial simply-connected entangling region with smooth boundary $\gamma_{\rm topo}=\ln \mathcal{D}$ where $\mathcal{D}$ is the effective quantum dimension of the underlying topological field theory,\cite{Kitaev2006,Levin2006} which is a topological invariant. In fact, in $d=2$ spatial dimensions, the von-Neumann EE and the R{\'e}nyi EEs have a rich structure since they also depend on the topology of the entangling regions and, for non-trivial topologies, $\gamma_{\rm topo}$ depends on the full structure of the topological field theory and not just on the effective quantum dimension.\cite{Dong-2008}
This  scaling law (and its generalizations) has been verified in many systems including fractional quantum Hall fluids,\cite{Zozulya-2009,Zozulya-2007,Haldane-2008}   topological phases of quantum dimer  models\cite{Furukawa-2007,Papanikolaou-2007,Stephan-2009} and  the related Kitaev's Toric Code model\cite{Hamma-2005,Hamma-2005b,Levin2006,Castelnovo-2008} (equivalent  to the $\mathbb{Z}_2$ gauge theory deep in its deconfined phase), and in chiral spin liquid phases of $d=2$ frustrated quantum antiferromagnets.\cite{Yi-2012} 

Much less is known about the scaling of EE in scale-invariant systems in $d>1$. Dimensional analysis and locality of the field theory suggest that that for scale-invariant systems in $d=2$ space dimensions with an entangling region with a smooth boundary, the EE again has the same form, $S_{vN}=\alpha \left(\frac{\ell}{a}\right)-\gamma$, where  the leading correction to the area law (perimeter in this case) is a finite  term $\gamma$.  
The finite term is expected to be scale-invariant  which, in general may be a universal function of the aspect ratios of the entangling region.\cite{Fradkin2006,Casini-2007}

This finite term has been computed explicitly in several cases but its general properties are not understood. In the case of the quantum Lifshitz model (QLM) it was computed by several authors.\cite{Hsu2009,Hsu2010,Stephan-2009,Stephan2011,Stephan-2012} The QLM is a scalar field theory in $d=2$ spatial dimensions with dynamical exponent $z=2$ (and hence not Lorentz invariant) which is the effective field theory of generalized quantum dimer models at their quantum critical points.\cite{Ardonne2004,Fradkin-2004}  Of particular interest to us is a result of Ref.[\onlinecite{Stephan2013b}] who gave a full  expression of the finite universal subleading term of the EE of the QLM for cylindrical entangling sections of a torus in the form of a scaling function of the aspect ratios of the cylinder. 

There has been great progress in understanding of scaling of the von Neumann EEs for entangling regions with the shape of a disk in $2+1$-dimensional relativistic conformal field theories.\cite{Casini-2007,Casini-2010} The result has  the same form as the EEs found in the QLM. In this context, in the literature the constant term is called $F$ (see, {\it e.g.} Ref.[\onlinecite{Klebanov2011}]). Casini {\it et al.}\cite{Casini-2011}  
have provided a proof in arbitrary dimensions of the holographic entanglement entropy ansatz of Ryu and Takayanagi\cite{Ryu2006} for the case of spherical entangling regions. In $2+1$ dimensions this result shows that the finite part of the entanglement entropy of a disk with a smooth boundary  is universal at a CFT. Additionally it was shown in \cite{Casini-2012ei} that, when appropriately defined, this finite part of the EE decreases under relevant perturbations of the CFT (and hence obeys a ``c-theorem''.)  Earlier results have given explicit values of $F$ for a disk for a free massless scalar field in $2+1$ dimensions.\cite{Casini-2010,Dowker-2010} In the case of the CFT of an interacting scalar field at its non-trivial Wilson-Fisher (IR) fixed point, it is  known for the case of a spatial split cylinder but only within the $1/N$ and $4-d=\epsilon$-expansions,\cite{Metlitski} and the extrapolation to $2+1$ dimensions is presently not understood. On the other hand, logarithmic contributions to the EE are found when the entangling region has cusp-like conical singularities,\cite{Casini-2007} are also found in the $z=2$ quantum Lifshitz model,\cite{Fradkin2006,zalatel-2011,Kallin-2014,Stoudenmire-2014} and at the quantum critical point of the ($z=1$) two-dimensional transverse field Ising model,\cite{Inglis2013} as well as in broken symmetry states with Goldstone bosons.\cite{Ding2008,Metlitski-2011,Ju2012}

Quantum Monte Carlo simulations have been used recently to compute the R{\'e}nyi entropy $S_2$  for several model wave functions of interest in condensed matter physics.\cite{Hastings-2010} St{\'e}phan and coworkers\cite{Stephan2013b} investigated the scaling of $S_2$ in cylindrical sections of a torus for the case of resonating-valence-bond (RVB) wave functions and for the wave functions of quantum dimer models on the square lattice. They also derived an explicit expression for the subleading term in the context of the QLM (which is believed to describe the continuum limit of these critical states), which is a universal scaling function of the aspect ratios of the cylinder. As expected, in the case of the quantum dimer model on the square lattice, the finite subleading term (for cylinders with aspect ratio 1)  extracted from their Monte Carlo results is clearly well fit by the universal scaling function deduced from the QLM.  In a separate study,\cite{Inglis2013} this group also investigated the scaling of the R{\'e}nyi entropy $S_2$ at the quantum critical point of the two-dimensional  Ising model in a transverse field. This system, which is in the same universality class as the classical three-dimensional Ising model, is Lorentz-Invariant at the quantum critical point, where it is described by an interacting  one-component relativistic real scalar field theory at its Wilson-Fisher (IR) fixed point. Remarkably, these authors find that the numerically obtained  R{\'e}nyi entropy $S_2$ is also well fitted (within a precision of a  fraction of 1\%) by the same scaling function derived from the QLM. This is quite unexpected since the QLM  has dynamical exponent $z=2$ and a global $U(1)$ symmetry whereas the quantum Ising model has a $\mathbb{Z}_2$ global symmetry and dynamical scaling exponent $z=1$ at the criticality. This apparent agreement is quite puzzling since these different universality classes are described by fixed points with very different scaling behaviors.

In this paper we re-examine the problem of the scaling of entanglement in two spatial dimensions using two different approaches. Firstly we consider a class of theories with relativistic critical points (CFTs)
that have the property that they are dual to a gravitational like theory in one higher dimensions, via the holographic
duality. In this case the Ryu-Takayanagi  ansatz can be used to  derive an explicit expression for the von Neumann EE for cylindrical sections of the torus by mapping the problem to a minimal surface computation in the anti-de Sitter (AdS) geometry (more precisely, we consider the AdS soliton geometry in order to have the torus topology on the boundary). Our result has a leading area law term and a finite sub-leading term which is a function of the aspect ratio of the cylindrical region that is being observed. We argue that in the ``thin slice'' limit, the pre-factor of the finite term is analogous to a central charge and is intrinsic to the $2+1$-dimensional CFT, giving a rough measure of the number of degrees of freedom in the theory. We will then use this ``central charge'' to rescale the finite sub-leading term,
thus allowing comparison of the functional dependence of the sub-leading term  across different theories.

Next we examined two simple free fermion field theories in $2+1$ dimensions where the different proposals for the scaling of entanglement can be tested directly. The first model is a theory of free Dirac fermions. In this case we used a lattice regularization in the form of spinless  fermions on a square lattice with flux $\pi$ per plaquette, which is a discretization of the Dirac fermion  known  as the Kogut-Susskind fermion.\cite{Kogut-1975} In two spatial dimensions, the low-energy limit of this model is equivalent to the two species (or ``valleys'')  of massless Dirac fermions with opposite parity,\cite{Fisher-1985}  analogous to  the case of graphene.\cite{Semenoff-1984} All local perturbations of this system are irrelevant operators and this is an infrared stable fixed point of the renormalization group. However, on a cylinder of finite radius this system behaves asymptotically as a system of free Dirac fermions in $1+1$ dimensions which is a CFT. 
 The second free fermion model we considered is a system  of fermions with two bands with a symmetry-protected quadratic band touching (QBT).\cite{Sun2009} In the low-energy limit, this system is equivalent to a theory of massless Dirac spinors with a quadratic dispersion and hence has the dynamical exponent $z=2$. In contrast to the massless Dirac fermion, this massless ``Lifshitz-Dirac'' fermion is an infrared unstable fixed point of the renormalization group and, in fact, all four-fermion operators are marginally relevant perturbations. Contrary to the case of free Dirac fermions, the QBT model on a cylinder of finite radius is not a $1+1$-dimensional CFT and has instead ultra-local correlations. Therefore the two fixed point theories have quite different dynamical properties. Since they are free-field theories, the EE can be computed explicitly with great accuracy\cite{Peschel2003} where the different proposals can be tested. 
 
The QBT model is also interesting in that it has a finite density of states (DOS) at low energies (while in the relativistic Dirac fermion case the DOS scales linearly with the energy). In this sense, the QBT model is reminiscent of the problem of fermions at finite density which has a finite DOS at the Fermi surface. In this case, it is known\cite{Wolf2006,Gioev2006,Swingle2010,Ding2012,Calabrese-2012,Leschke-2014} that the von Neumann EE has a logarithmic violation of the area law 
 of the form $S_E=\alpha (\frac{l_A}{\epsilon})^{d-1}\log\frac{l_A}{\epsilon}$, where the prefactor $\alpha$ has been argued to be essentially universal provided the scale $\epsilon$ is determined  by the size of the Fermi surface (see, however, the  numerical results  of Ref.[\onlinecite{McMinis2013}]). 
This result may suggest that the finite DOS of a Fermi liquid at the Fermi surface may be the origin of the logarithmic violation of the area law, and that systems with a finite DOS at asymptotically low energies may also obey a similar scaling law. We will see, however, here that  this is not the case.

Keeping the differences in mind, we studied the two-cylinder EEs of both fermionic models by computing the EEs of the cylinder explicitly (albeit numerically). In spite of  the  differences in physics, we find that the EEs of the models satisfy the area law and, in particular in the case of the QBT,  we do not find  any logarithmic violation of entanglement scaling from the area law.  We further study the scaling behavior of the subleading term in the EEs.
In the case of massless Dirac fermions we find that although the expression derived from the QLM fits well with surprising accuracy, the holographic entropy result for the cylinder appears to be essentially exact. In the case of the QBT the finite subleading term in the EE is accurately fitted by the expression derived from the QLM. 

The rest of this paper is organized as following. In the section II, we introduce and explain the three possible scaling functions which will be tested in two free fermion models, namely a free Dirac fermion model and a QBT model. In the section III, we will explain, based on the asymptotic behaviors of equal-time two-point correlators, why the EE of the QBT cannot have any logarithmic violation of the area law. In the section IV, we numerically calculate the EE of the two fermion models and test the three scaling functions proposed in the section II. In the section V, we summarize our results and 
conclude that there is a universal scaling function of the subleading term of the EEs for the critical systems.  

\section{Entanglement Entropy Scaling Functions}
\label{sec_2}
In this paper we will discuss three possible EE scaling functions for scale-invariant systems in $d=2$ space dimensions. We will restrict ourselves to the EE of two cylinders A and B obtained from a partition of a torus. The  scaling functions enter as scale-invariant finite corrections of the leading, area law, term of the von Neumann and R{\'e}nyi entropies. They are: a) a quasi-1D scaling function, b) the quantum Lifshitz model scaling function, and c) a holographic scaling function (which we derive here using the AdS/CFT correspondence).  

\begin{figure}[th]
\centering
\includegraphics[scale=.4]{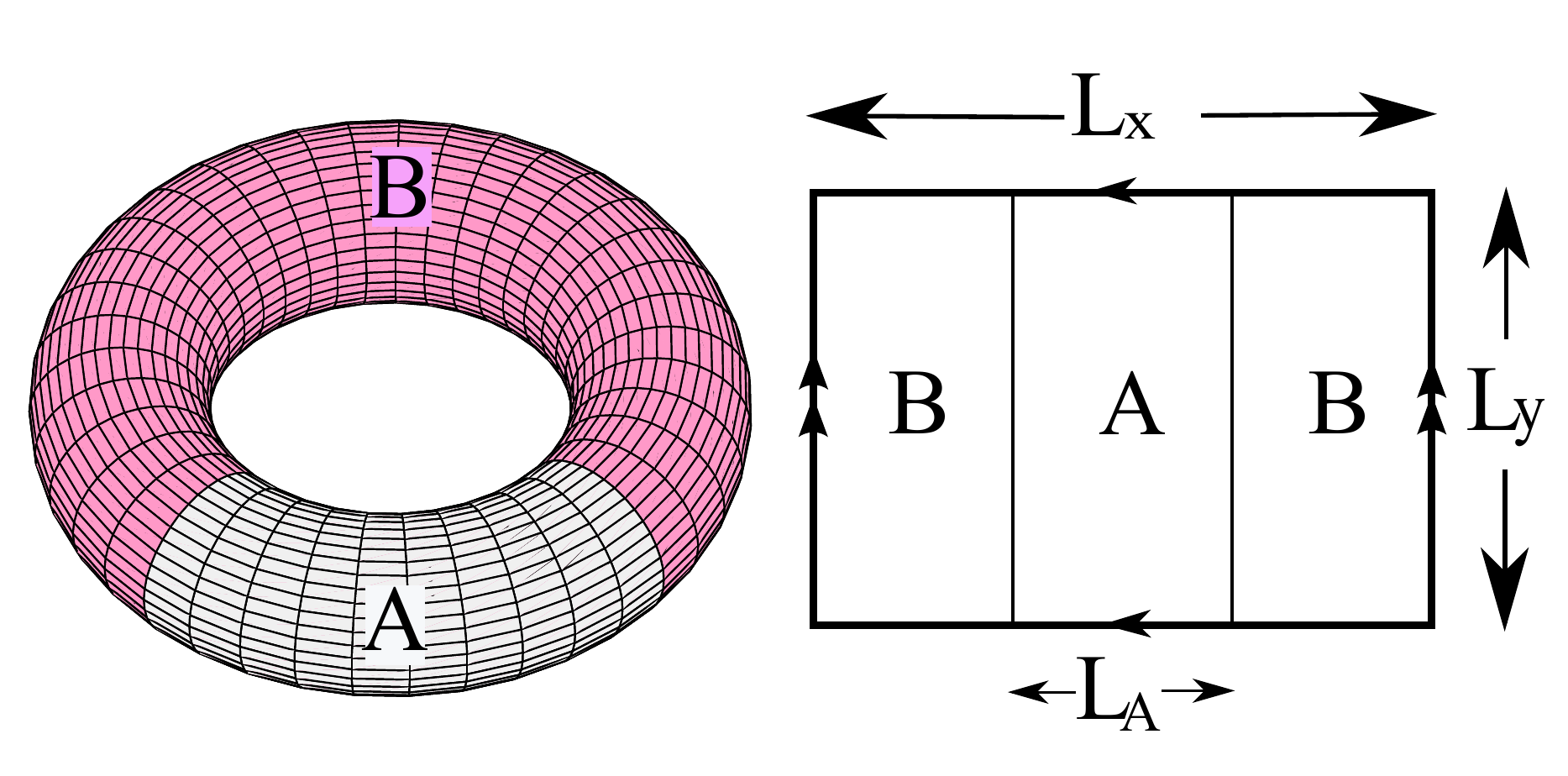}
\caption{The torus is divided into two cylinders A and B with size $L_A\times L_y$ and $(L_x-L_A)\times L_y$.  }
\label{fig:schematic}
\end{figure}

Different geometries of bipartition may give rise to different subleading terms with different structure. For instance, both numerical and analytical calculations on $2+1$-dimensional critical models show that there is a subleading term correction if the boundary of the subregion A is not smooth.\cite{Stephan2011, Kallin-2014} The corner will give rise to the logarithmic term in the EE with the coefficient proportional to the low-energy degrees of freedom.\cite{Inglis2013, Kallin-2014, Stoudenmire-2014} Even for the smooth boundary, the curvature on the subregion A may also lead to the logarithmic correction.\cite{Fradkin2006} In this paper, to avoid both the corner and the curvature corrections, we consider the torus geometry and bipartition the torus into two cylinders with a smooth boundary and calculate the two-cylinder entropy as shown in Fig. \ref{fig:schematic}. 

\subsection{Quasi-1D Entanglement Scaling Function}

This  scaling function was introduced heuristically by Ju and coworkers.\cite{Ju2012} It assumes that in the thin torus limit, $L_y \ll L_x$ (see Fig.\ref{fig:schematic}), the effectively quasi-one-dimensional system should approximate a $1+1$-dimensional CFT. The posited form of the von Neumann EE is\cite{Ju2012} (up to a non-universal additive constant)
\begin{equation}
S_{vN}=\alpha L_y+\beta \log \sin (\pi \y)
\end{equation}
where  $\y=L_A/L_x$. Here $\alpha$ is a non-universal coefficient and $\beta$ is universal. 

\subsection{Quantum Lifshitz Entanglement Scaling Function}

This scaling function was  derived from the QLM by Stephan and coworkers,\cite{Stephan2013b} who tested it in the quantum dimer model (on the square lattice) and in the two-dimensional Ising model in a transverse field.\cite{Inglis2013} For a torus with aspect ratio $L_y/L_x$, the von Neumann EE is
\begin{equation}
S_{vN}=\alpha L_y+ \beta J(\y)
\label{ad}
\end{equation}
where  $J(\y)$ is given by\cite{Stephan2013b}
\begin{align}
J(\y)= 
\log \left(\frac{\lambda}{2}  \frac{\eta(\tau)^2}{\theta_3(\lambda \tau)\theta_3(\tau/\lambda)}  \frac{\theta_3(\lambda \y\tau)\theta_3(\lambda (1-\y)\tau)}{\eta(2\y\tau)\eta(2(1-\y)\tau)}\right) 
\label{j_func}
\end{align}
where $\theta_3(z)$ is the Jacobi theta-function, $\eta(z)$ is the Dedekind eta-function, $\tau=i L_x/L_y$ is the modulus of the torus, and $\lambda$ is a parameter. For the case of the quantum dimer model at its Rokhsar-Kivelson quantum critical point the parameter is $\lambda=2$. In this paper we will test this scaling function in two free fermion models in $2+1$-dimensions and use $\lambda$ as a fitting parameter. As we will see this scaling function works surprisingly well even in relativistic systems.

\subsection{Holographic Relativistic Entanglement Entropy}

To get a handle on the surprising universality of the scaling function $J(\y)$ we now turn to another set
of quantum systems whose EE can be efficiently calculated, {\it i.e.}, strongly-interacting relativistic (with dynamical exponent $z=1$)
quantum field theories which are described by a weakly-coupled dual gravity theory. 
There is a large class of such examples and we will concentrate on a subset which can be effectively described by (rather, truncated to) AdS gravity in $3+1$ dimensions. 
Since $J(\y)$ is defined on a torus geometry, we must pick the appropriate solution to Einstein's equations with torus boundary topology (in the spatial directions). This is the AdS soliton 
metric.\cite{Witten-1998, Horowitz-1998}  There are actually two possible metrics that we can use, depending on which torus direction
($x$ or $y$) we allow to contract in the bulk - picking the smallest direction describes the ground state of the system. 

The EE is sensitive to which cycle of the torus contracts because the cut is always along the $y$-direction. We will study both cases in detail. In the case where $L_y < L_x$ and the $y$-cycle contracts one finds that the EE  saturates for large enough $L_A$ (but still smaller compared to $L_x$) - this can be understood by taking the thin torus limit $L_y \ll L_x$ where it is clear that the saturation indicates the effective low energy $1+1$-dimensional theory is gapped. 
The reason for this can be traced to the anti-periodic boundary conditions for fermions around the torus cycles,  which is forced upon us just by the fact that we allow such gravitational solutions with contracting spatial cycles \cite{Witten-1998}. Periodic boundary conditions could also be studied, however this presumably would involve more stringy ingredients (for example the application
of T-duality to the contracting cycle) and the calculation of EE in such situations is not developed.

We will eventually compare the strongly interacting holographic model to the free Dirac model with periodic boundary conditions at $L_x=L_y$ and so not surprisingly the geometry where the $y$-cycle contracts does not do a good job due to this saturation. However it turns out that the phase where the $x$-cycle contracts, which is not continuously connected to the phase showing the aforementioned gap, has an incredibly similar form to the Dirac answer. We consider this case taking $L_y \geq L_x^+$
and return to the other case later.

\subsubsection{AdS soliton geometry with $L_y\geq L_x$}

According to the Ryu-Takayanagi conjecture,\cite{Ryu2006} the EE takes the very simple form:
\begin{equation}
S=\frac{\mathcal{A}}{4G_N}
\label{HolographicEE}
\end{equation}
where $\mathcal{A}$ is the Area of the minimal surface ending on the boundary where the
QFT lives at $\partial \mathcal{A}$ and falling into the bulk AdS-soliton geometry. There is by now ample evidence for this formula \cite{Lewkowycz:2013nqa} 
and so we will take it as a given.

The AdS soliton metric is given by:
\begin{equation}
ds^2=\frac{1}{z^2}\left(\frac{dz^2}{f}+fdx^2+dy^2-dt^2\right)
\end{equation}
where $f = 1-(z/z_h)^3$. This geometry looks like a cigar in the $(x,z)$ directions, where the tip is at $z=z_h$ and $x$
is the angular direction. To avoid the conical singularity at the tip, we need to impose the constraint: $x\sim x+\frac{4\pi}{3}z_h$. Since $x$ has the periodicity $x\sim x+L_x$, we require $z_h=\frac{3}{4\pi}L_x$. 

The minimal surface for the subregion $\mathcal{A}$ can be calculated by assuming an ansatz
which is translationally invariant in the $y$ direction and has profile: $x(z)$. 
\begin{eqnarray}
\nonumber \mathcal{A}&=&\int \sqrt{G}dzdy=\int dzdy\left[\frac{1}{f}+f(x^{\prime})^2\right]^{1/2}\frac{1}{z^2}\\
\nonumber &=&2L_y\int_{\epsilon}^{z_\star}dz\left[\frac{1}{f}+f(x^{\prime})^2\right]^{1/2}\frac{1}{z^2}\\
&=&2L_y\int_{\epsilon}^{z_\star}L(x,x^{\prime},z)
\end{eqnarray}
where $x^{\prime}=dx/dz$ and $G$ is the induced metric 
on the co-dimension $2$ surface $(t=0,x = x(z) )$.

The minimal area profile can be found using standard Lagrangian mechanics: $\delta L/\delta x=0$
from which the equation of motion is $E=\frac{\partial L}{\partial x^{\prime}}=x^{\prime}f\left[1/f+f(x^{\prime})^2\right]^{-1/2}/z^2$.
This leads to
\begin{equation}
(x^{\prime})^2=\frac{1}{f^2}\frac{E^2}{f/z^4-E^2}
\end{equation} 
We have defined the point $z=z_\star$ such that $x^{\prime}=\infty$ which will be
the largest $z$ obtained by the surface. $z_\star$ satisfies $f_\star/z_\star^4=E^2$, where $f_\star=1-(z_\star/z_h)^3$.

Integrating the above differential equation we can solve for $z_\star$ in terms of $L_A$:
\begin{eqnarray}
\nonumber \frac{L_A}{2}&=&\int_{-L_A/2}^0 dx=\int_0^{z_\star}\frac{E}{f}\left(\frac{1}{f/z^4-E^2}\right)^{1/2}dz\\
\nonumber &=&z_\star\int_0^1 d\zeta \frac{1}{f}\left[\frac{1}{(\frac{f}{f_\star})(\frac{1}{\zeta})^4-1}\right]^{1/2}
\\
&=&\frac{L_x}{2}  \y( \chi )
\label{constraint}
\end{eqnarray}
where $\zeta=z/z_\star$ and $\chi$ is related to the
turning radius $z_\star$ of the minimal surface $\chi = (z_\star/z_h)^3$. The final form of $\y$
is:
\begin{align}
\label{sfa}
\y (\chi) &= \frac{3 \chi^{1/3} (1-\chi)^{1/2}}{2 \pi} \int_0^1 \frac{ d \zeta \zeta^2}{ (1- \chi \zeta^3)} \frac{1}
{ \sqrt{P(\chi,\zeta)} } 
\end{align}
where $P(\chi,\zeta) = 1- \chi \zeta^3 - (1-\chi) \zeta^4$. 

By solving the above equation, we can obtain  $\y=L_A/L_x$ for different values of $\chi$.

The area of the minimal surface equals to
\begin{eqnarray}
\nonumber  \mathcal{A} &=&2L_y\int_{\epsilon}^{z_\star}dz\frac{1}{z^4}\left(\frac{1}{f/z^4-E^2}\right)^{1/2}\\
\nonumber &=&\frac{2L_y}{z_\star}\int_{\epsilon/z_\star}^1d \zeta\frac{1}{(f_\star)^{1/2}\zeta^4}\left[\frac{1}{(\frac{f}{f_\star})\frac{1}{\zeta^4}-1}\right]^{1/2}\\
&=&\frac{2L_y}{\epsilon}+\frac{8 \pi L_y}{3 L_x} j(\chi)
\label{min_surface}
\end{eqnarray}
where  we have separated out the linearly divergent term, regulated by a cutoff close to boundary
at $z=\epsilon$. The first term in $\mathcal{A}$ is the divergent area law and the second term is the finite subleading correction which   
can be calculated numerically using a parametric description for $L_A$ and $j$ in terms of $0<\chi<1$. The final
form of $j$ is:
\begin{align}
j(\chi) &= \chi^{-1/3} \left(   \int_0^1 \frac{ d\zeta}{\zeta^2} \left( \frac{1}{ \sqrt{P(\chi,\zeta)}}  -1  \right)  -1 \right)
\label{sfj}
\end{align}

When $\y$ is small, $z_\star<<z_h$ and in this case, the metric is the same as the metric for the usual AdS space and the subleading term $j(\chi)$ is proportional to $1/\y$ which was first calculated in Ref. [\onlinecite{Ryu2006}].

In order to compare with the Dirac model we should normalize the coefficients in front of $1/\y$ 
to be the same in the two cases. This requires some explanation - we are working in the classical
gravity limit where $G_N \rightarrow 0$, so for the results we quoted to hold the coefficient
in front of $1/\y$ will be very large. This is certainly not the case for the Dirac model. In order
to effectively compare these results we should then take a large number of copies of the Dirac model, with no interactions amongst each copy. The EE for the Dirac model then scales accordingly and in this way we can have a large $1/\y$ coefficient to compare to the holographic model.

For comparison to the QBT model, a better holographic dual model will have a different metric (related to the $z=2$ Lifshitz space-times introduced in \cite{Kachru:2008yh}). It is not hard to see that when $\y$ is small, it should have the same scaling behavior as the Dirac model. We leave comparison of the subheading terms in the $z=2$ case to future work.

\subsubsection{AdS soliton geometry with $L_y\leq L_x$}

Similar expressions may be derived for the case where $L_y < L_x$. In this case
the situation is complicated by the existence of a disconnected minimal surface that
fills in the contractible $L_y$ cycle of the AdS-soliton. This causes a saturation in the EE
which we interpret as a gap for the lower dimensional system after a low energy reduction along the $y$
direction. This saturation is related to the phase transition in holographic EE studied in Ref.
[\onlinecite{Nishioka:2006gr}].
The appropriate scaling form is:
\begin{align}
\mathcal{A}- \frac{2 L_y}{\epsilon}  & = \frac{ 8 \pi}{3} \widetilde{j}\left( \frac{L_x}{L_y} \y \right) 
\,,  \quad 0 < \y < \frac{L_y}{L_x}  p  \\
 & = - \frac{8 \pi}{3}\,,     \quad \qquad \frac{L_y}{L_x} p < \y < 1 - p  \frac{L_y}{L_x} \\
 & = \frac{ 8 \pi}{3} \widetilde{j}\left( \frac{L_x}{L_y} (1-\y) \right) 
\,, \quad 1  - p  \frac{L_y}{L_x} <  \y < 1 
\end{align}
where $p \approx 0.19$ is a fixed number determined by where the saturation of EE occurs (the middle equation above).  The function $\widetilde{j}$ is defined parametrically:

\begin{align}
\widetilde{j}(\chi) & = \chi^{-1/3} \left( \int_0^1 \frac{d \zeta}{\zeta^2} \left( \frac{\sqrt{1 - \chi \zeta^3}}{\sqrt{P(\chi,\zeta)}} -1 \right) -1 \right)  \\
\frac{L_x}{L_y} \y & = \frac{3}{2\pi} \chi^{1/3} (1 - \chi)^{1/2} \int_0^1 \frac{d\zeta \zeta^2}{\sqrt{1-\chi \zeta^3}} \frac{1}{\sqrt{P(\chi,\zeta)}}
\end{align}

\begin{figure}[hbt]
\centering
\includegraphics[scale=.33]{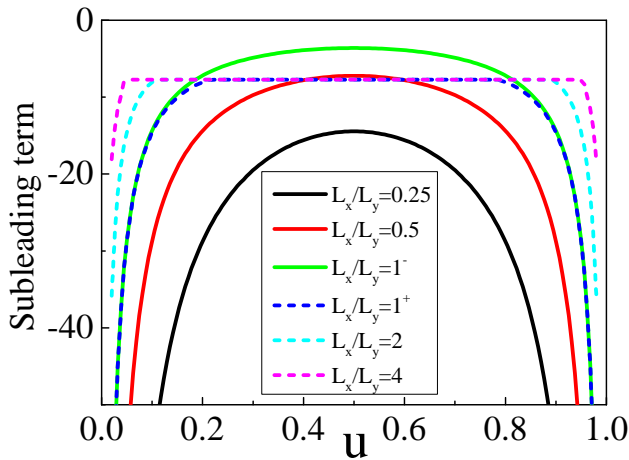}
\caption{The subleading term for the minimal surface for various values of $L_x/L_y$. The solid curves
are for $j(\y)$ when $L_x \leq L_y$ and the dashed curves are for $\widetilde{j}(\y)$ when $L_x > L_y$. See text for details.}
\label{fig:fig8}
\end{figure}

For completeness we plot the full set of scaling forms of $j(\chi)$ and $\widetilde{j}(\chi)$ for different values of $L_x/L_y$
in Fig. \ref{fig:fig8}. 

\begin{figure}[hbt]
\centering
\includegraphics[scale=.33]{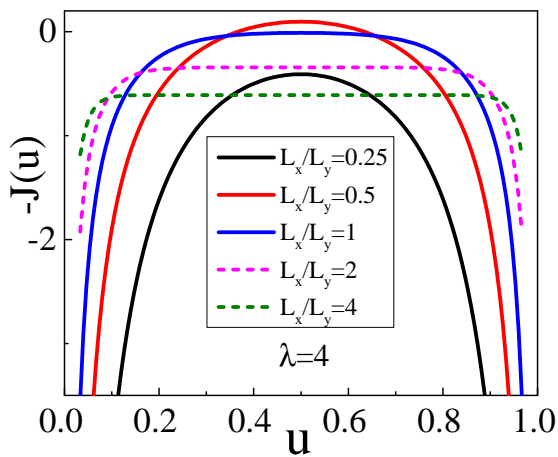}
\caption{$-J(\y)$ for various values of $L_x/L_y$. The solid curves
are for $L_x \leq L_y$ and the dashed curves are for $L_x > L_y$. }
\label{fig:fig9}
\end{figure}

We also plot the complete scaling forms of $J(\y)$ with different aspect ratio $L_x/L_y$ in Fig. \ref{fig:fig9}. The $J(\y)$ is defined in Eq.~(\ref{j_func}). \cite{Stephan2013}
As shown in Fig. \ref{fig:fig9}, the $J(\y)$  function has similar scaling behavior as the $j(\y)$ and $\widetilde{j}(\y)$ function. In the thin torus limit, $J(\y)$ also shows the saturation behavior around $\y=0.5$. 

\section{Area law for QBT model}

The fermionic QBT model in $2+1$-dimensions is a free fermionic spinor model with a quadratic energy dispersion. The Hamiltonian of the model has a similar structure as the Dirac fermion,
\begin{equation}
H=\int \frac{d^2k}{(2\pi)^2} \Psi^{\dagger} (\bm k) \begin{pmatrix} k_x^2-k_y^2 & -2ik_xk_y \\ 2ik_xk_y& -k_x^2+k_y^2 \end{pmatrix}\Psi(\bm k), 
\label{QBT:1}
\end{equation}
with $\Psi(\bm k) = \begin{pmatrix} \psi_1(\bm k) & \psi_2(\bm k) \end{pmatrix}^{T}$. The energy spectrum for this model is $E(\bm k)=\pm (k_x^2+k_y^2)$. Different from a Dirac model, any local four-fermion term is marginally relevant,\cite{Sun2009} which means that even an infinitesimally weak interaction leads to an instability of the free QBT point to the spontaneous breaking of either time-reversal invariance or the point group symmetry of the lattice. The QBT model, in this sense is a critical point \cite{Sun2009}. In contrast, the Dirac model in $2+1$ dimensions is a stable fixed point since all local  interactions are irrelevant at low energies. For the QBT model, at the band touching point $\bm k=(0,0)$, there is a finite DOS. Since the origin of the violation of area law for the EE in a Fermi liquid is Fermi surface (which has a finite DOS), one might speculate that the QBT model may be a ``Fermi liquid" of sorts with the Fermi surface replaced by a Fermi point and that it would also break the area law. To see if there is any violation of the area law in the EE of the QBT model, we first study the two-point equal-time correlation function for the fermionic QBT model. For the free fermion system, the entanglement entropy can be obtained by calculating the two-point correlation function, and hence the correlation function in the long distance limit can give information about the EE at the thermodynamic limit.

The Lagrangian density for the QBT model is
\begin{equation}
{\mathcal L}=\bar{\Psi}\left[i\gamma_0\partial_0-i(\partial_1^2-\partial_2^2)\gamma_1+2i\partial_1\partial_2\gamma_2\right]\Psi.
\end{equation}
where $\gamma_0=\sigma_1$, $\gamma_1=\gamma_0\sigma_3$ and $\gamma_2=\gamma_0\sigma_2$ and $\bar{\Psi} = \Psi^{\dagger}\gamma_0$.

To calculate the correlation function of the QBT model, we first calculate the equal-time correlation function for the $2+1$-dimensional QLM.\cite{Ardonne2004} The equal-time two-point correlation function of the QLM has the asymptotic behavior in $|{\bm r}| = \sqrt{x^2 + y^2}\to\infty$ with $|{\bm r}|$ as the spatial distance between the two points:
\begin{equation}
G_{\rm QLM}(\bm r) = \frac{1}{4\pi } \log(|{\bm r}|).
\end{equation}
From this, we obtain the two point correlation function for the QBT model
\begin{eqnarray}
\nonumber G_{\rm QBT}(\bm r)&=& ((\partial_1^2-\partial_2^2)\gamma_1-2\partial_1\partial_2\gamma_2)G_{\rm QLM}(\bm r)\\
&=&-\frac{2(x^2-y^2)\gamma_1-4xy\gamma_2}{4\pi r^4}.
\end{eqnarray}

On the other hand, we can also calculate the correlation function for the Dirac fermion from the bosonic model. We calculate the equal-time correlation function of the relativistic free massless scalar field. In the limit $r\to \infty$:
\begin{equation}
G_{0}(r)\propto \frac{1}{|r|}.
\end{equation}
Hence the two point correlation function for the Dirac fermion at equal time is 
\begin{eqnarray}
\nonumber G_{\rm D}(\bm r)&=& (\gamma_i\partial_i)G_{0}(\bm r)\\
&\propto&-\frac{x\gamma_1+y\gamma_2}{r^3}.
\end{eqnarray}

For the QBT and Dirac models, we can see that the two-point correlation functions have asymptotically identical behavior in the long distance limit $|\bm r| \to \infty$. This implies that both the models will have the same scaling behavior for the leading term of the EE when the sizes of the subsystem is large enough compared to the UV cutoff. Since the Dirac model obeys the  area law,\cite{Ryu2006} the QBT model should also obey the area law and cannot have more divergent terms in the EE than the area law allows, in spite of having a finite DOS at zero energy.

Since the QBT model satisfies the area law, it is less entangled than  the Fermi liquid. On the other hand, since the QBT model is a scale-invariant system with an IR unstable fixed point we expect it to have long-range entanglement in the form of scale-invariant contributions to the EE, which can only enter in the form of an $O(1)$ finite subleading correction to the area law.
However, the correlation function argument itself cannot tell much information about the structure of the subleading term. To study the subleading term in EE for the QBT model we will need an explicit expression. Unfortunately it is not possible to write the EE as a closed analytic expression and we will use instead numerical methods to study its scaling behavior.

\section{Entanglement entropy for  Dirac and QBT fermions}

The lattice model of the QBT model  in momentum space can be written as 
\begin{equation}
H_{\rm QBT}=\int_{BZ} \frac{d^2k}{(2\pi)^2}\Psi^{\dag}(\bm k)\mathcal{H}_{\rm QBT}(\bm k) \Psi(\bm k),
\label{Lif_spe}
\end{equation}
where $BZ$ stands for the first Brillouin zone, $-\pi< k_x\leq\pi$ and $-\pi< k_y\leq\pi$. Here $\Psi^{\dag}(\bm k)$ is a two component spinor fermionic creation operator $\Psi^{\dag}(\bm k)=(\psi^{\dag}_1(\bm k),\psi^{\dag}_2(\bm k))$. The one-particle Hamiltonian $\mathcal{H}_{\rm QBT} (\bm k)$ has the form
\begin{equation}
\mathcal{H}_{\rm QBT}(\bm k)=h_1(\bm k)\sigma_1+h_3(\bm k)\sigma_{3}, 
\end{equation}
where  $\sigma_1$ and $\sigma_3$ are the usual two Pauli matrices, and 
\begin{align}
h_1(\bm k)=&-4t\cos(\frac{k_x}{2})\cos(\frac{k_y}{2}),\nonumber \\
h_3(\bm k)=&-t^{\prime}(\cos(k_x)-\cos(k_y)). 
\end{align}
The QBT point is at $\bm k=(\pi,\pi)$. Near the point, we can expand $\mathcal{H}_{\rm QBT}$ and find the continuum Hamiltonian Eq.\eqref{QBT:1}.  In the numerical calculation, we will set $t=t^{\prime}=1$. 

Similarly, the lattice model for the Dirac fermion is a tight-binding model of spinless fermions on the square lattice with $\pi$ flux on each plaquette. The Hamiltonian in momentum space is
\begin{equation}
H_D=\int_{BZ} \frac{d^2k}{(2\pi)^2} \Psi^{\dag}(\bm k)\mathcal{H}_D(\bm k) \Psi(\bm k),
\end{equation}
where $BZ$ stands for the first Brillouin zone, $-\pi< k_x\leq\pi$ and $-\pi< k_y\leq\pi$. The one-particle lattice Dirac Hamiltonian $\mathcal{H}_D(\textbf{k})$ takes the form
\begin{equation}
\mathcal{H}_D({\bm k})=h_1\sigma_1+h_3\sigma_{3},
\end{equation}
with  $h_1=-2\cos(k_x)$ and $h_3=2\cos(\frac{k_y}{2})$. The Dirac points are at $(\pm \frac{\pi}{2},\pi)$. 

In the numerical calculation, we put these two models on the torus as shown in Fig.~\ref{fig:schematic} and calculate the two-cylinder entropy when the lower band is fully filled. In this geometry, the momentum $k_y$ parallel to the cut is always a good quantum number, thus the $2+1$-dimensional model can be considered as a set of $1+1$-dimensional chains with an effective mass depending on the value of $k_y$. Thus we calculate the total two dimensional EE as the sum over the EE of the $1+1$-dimensional chains labelled by the momentum $k_y$. Furthermore, we notice the R{\'e}nyi entropies $S_n$ with different R{\'e}nyi indices $n=1,2\ldots$ show similar behavior and we will only consider the von Neumann entropy $S_{vN}$ later in this paper. By analogy with the QLM\cite{Stephan2011} , one might worry that there may be a phase transition between $n=1$ and $n=2$, but we find there is none in that the R{\'e}nyi entropies $S_n$ with different R{\'e}nyi indices $n=1,2\ldots$ can be fitted with the area law term supplemented by a single universal scaling form for the subleading term.

\begin{figure}[th]
\centering
\includegraphics[scale=.35]{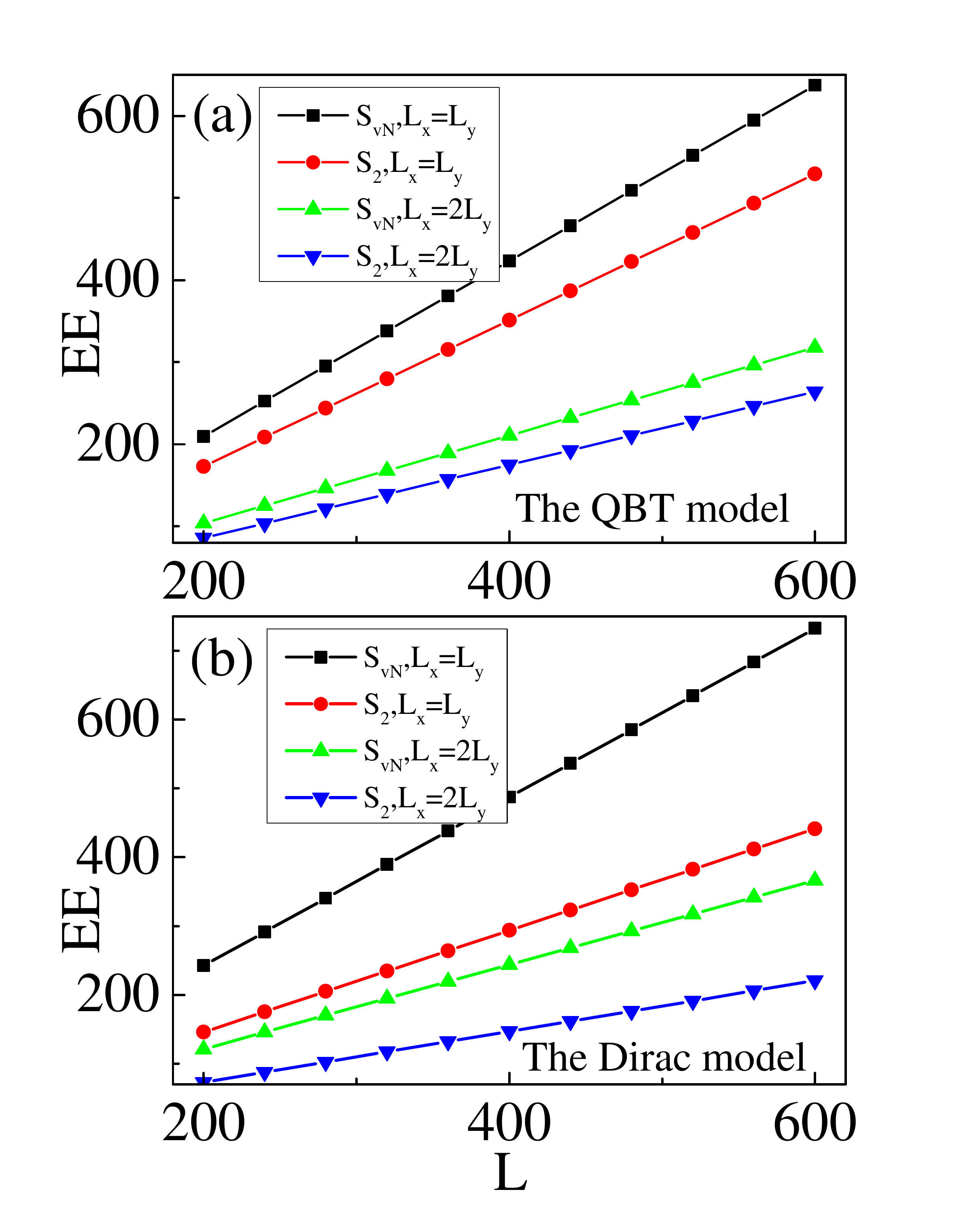}
\caption{(a) The von Neumann entropy $S_{vN}$ and R{\'e}nyi entropy $S_2$ for the QBT model as a function of $L=L_x$. Both the EEs are linearly proportional to $L$ when the aspect ratio $L_x/L_y$ and the ratio $\y=L_A/L_x$ are fixed. $\y=0.1$ and two different aspect ratios are considered. (b) The von Neumann entropy $S_{vN}$ and R{\'e}nyi entropy $S_2$ for the Dirac model. The setup of the bipartition geometry is the same as (a).}
\label{fig:area_law}
\end{figure}

We first check that both models satisfy the area law numerically. 
To verify that the EE satisfies the area law, we change the length of $L_y$ but fix the aspect ratio $L_x/L_y$ to be a constant value. The numerical calculation for both the models are shown in Fig.~\ref{fig:area_law}(a) and (b). For the QBT and Dirac models, the von Neumann entropy and R{\'e}nyi Entropy with R{\'e}nyi index $n=2$ are all linearly proportional to $L_y$, $S_E=\alpha L_y+\gamma$. The coefficient in front of $L_y$ is invariant when we change the ratio $\y=L_A/L_x$. This indicates that both models satisfy the area law.

\begin{figure}[th]
\centering
\includegraphics[scale=.34]{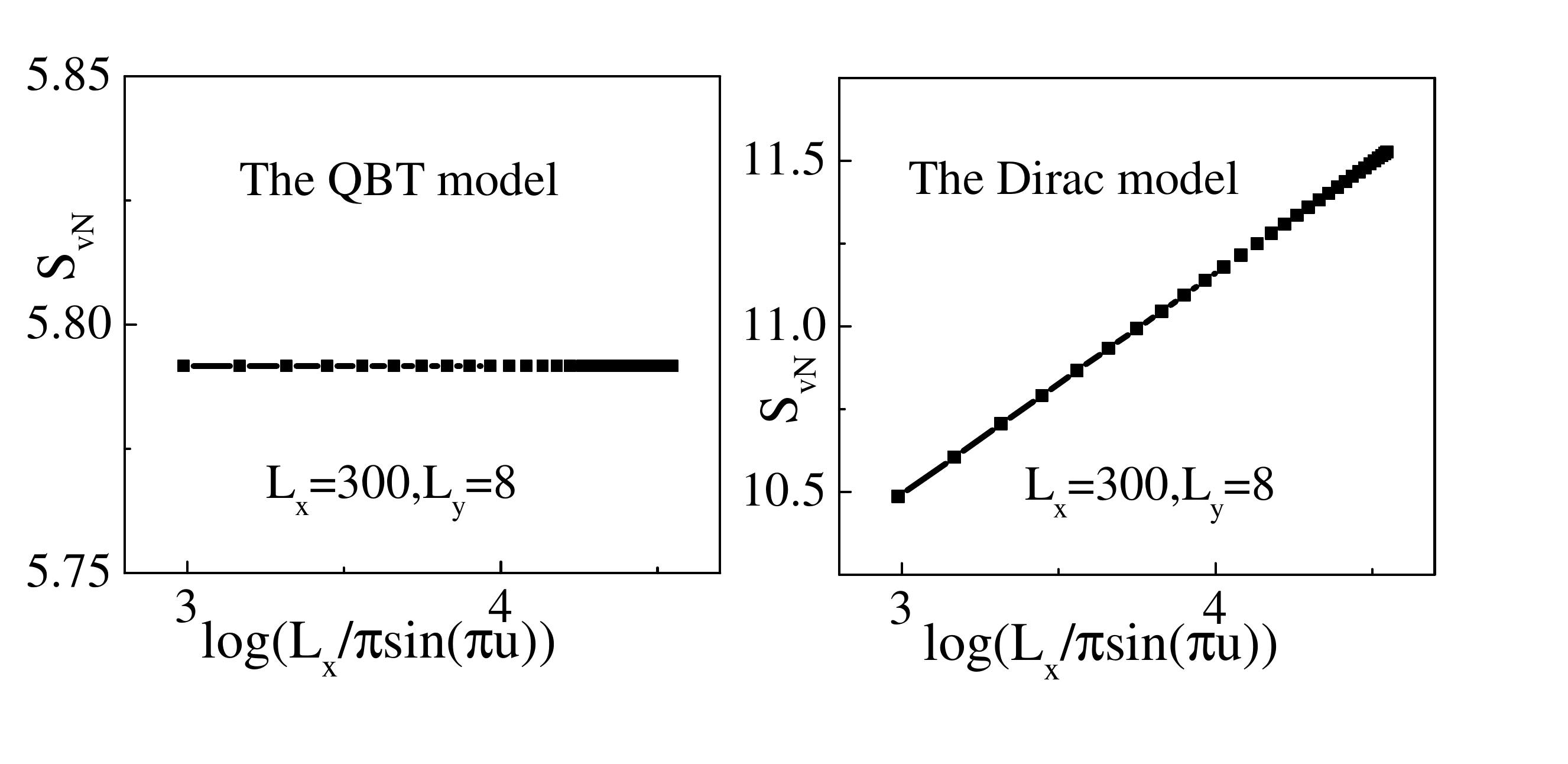}
\caption{Left: The von Neumann entropy $S_{vN}$ for the QBT model in the thin torus limit. Right: $S_{vN}$ for the Dirac model in the thin torus limit with periodic boundary condition in $y$ direction. The coefficient is $0.666$ since there are two Dirac cone in the lattice model.}
\label{fig:thin_torus}
\end{figure}

We will compute the finite subleading term and consider two different regimes for the aspect ratio $L_x/L_y$ of the torus: the thin torus limit $L_x<<L_y$ and the two dimensional limit $L_x\approx L_y$. In the thin torus limit, the models are expected to behave effectively as the $1+1$-dimensional theory. For the Dirac model, the computation  is sensitive to the boundary condition in the $y$ direction (Fig.~\ref{fig:schematic}). For  periodic boundary conditions, the zero mode $k_y=0$ will contribute the logarithmic correction $\frac{1}{3}\log L_A$ to the total EE,\cite{Calabrese2004} while for anti-periodic boundary conditions there is no zero mode and no such logarithmic correction. In contrast,  the QBT model is insensitive to the boundary condition. The zero mode $k_y=0$ contributes nothing to the total EE (it contributes only finite $O((L_y)^{0}) = O(1)$ contribution to the total EE). This is because in $1+1$ dimensions the  QBT model is not critical and is only short-ranged correlated. The two-point equal-time correlation function of the $1+1$ -dimensional QBT model is a delta-function. In particular, this also implies that here is no logarithmic subleading correction for the $2+1$-dimensional QBT model either. Thus, in the thin torus limit with periodic boundary condition, the Dirac model has a logarithmic subleading term correction while the QBT model does not. This result is verified by the numerical calculations shown in Fig.~\ref{fig:thin_torus}.

The above thin torus limit argument for the subleading term does not work in the two dimensional limit $L_x\approx L_y$ because of the complicated crossover behavior for the $1+1$-dimensional  massive Dirac fermion with mass $k_y\approx L_A^{-1}$, which becomes analytically intractable. Instead, we directly calculate the subleading term numerically and fit the data with possible candidates for subleading terms. Since we are considering free fermion system, we will use Peschel's result to calculate the EEs.\cite{Peschel2003}

The numerical results in the two dimensional limit show that the subleading term only depends on the aspect ratio of the torus $L_x/L_y$ and on the  ratio $\y=L_A/L_x$.\cite{Ju2012} For simplicity, we will fix $L_x/L_y=1$ (a square torus, and hence with modulus $\tau=i$) and only study the dependence of the subleading term on the aspect ratio $\y$. We test three possible scaling functions for the subleading term defined in Sec.\ref{sec_2}, 
\begin{eqnarray}
\label{J_fit}
S_{vN}&=&\alpha L_y+\beta J(\y)\\
\label{j_fit}
S_{vN}&=&\alpha L_y+\beta j(\y)\\
S_{vN}&=&\alpha L_y+\beta \log(\sin(\pi  \y))
\end{eqnarray}
where $J(\y)$ in Eq.(\ref{J_fit}) is given by\cite{Stephan2013b} 
\begin{equation}
\label{sf}
J(\y)=\log \Big(\frac{\theta_3(i\lambda \y)\theta_3(i \lambda(1-\y))}{\eta(2i\y)\eta(2i(1-\y))}\Big), 
\end{equation}
which is obtained from Eq.~\eqref{ad} by plugging $\tau = i$ and truncating $\log (\frac{\lambda \cdot \eta(i)^2}{2\theta_3(\lambda i)\theta_3(i/\lambda)})$, which is an O($1$) constant. This scaling function was originally derived from the QLM. However, it was also found  unexpectedly to fit well with the numerical results of the EE of the relativistic scalar field theories.\cite{Stephan2013,Inglis2013} $j(\y)$ in Eq.~(\ref{j_fit}) is derived from the holographic calculation shown in Eq.~(\ref{sfj}), we will only check it for the Dirac fermion model. We also consider $\log(\sin(\pi\y))$ as the possible subleading term because it is a natural extension of the thin torus limit.

\begin{figure}[hbt]
\centering
\includegraphics[scale=.37]{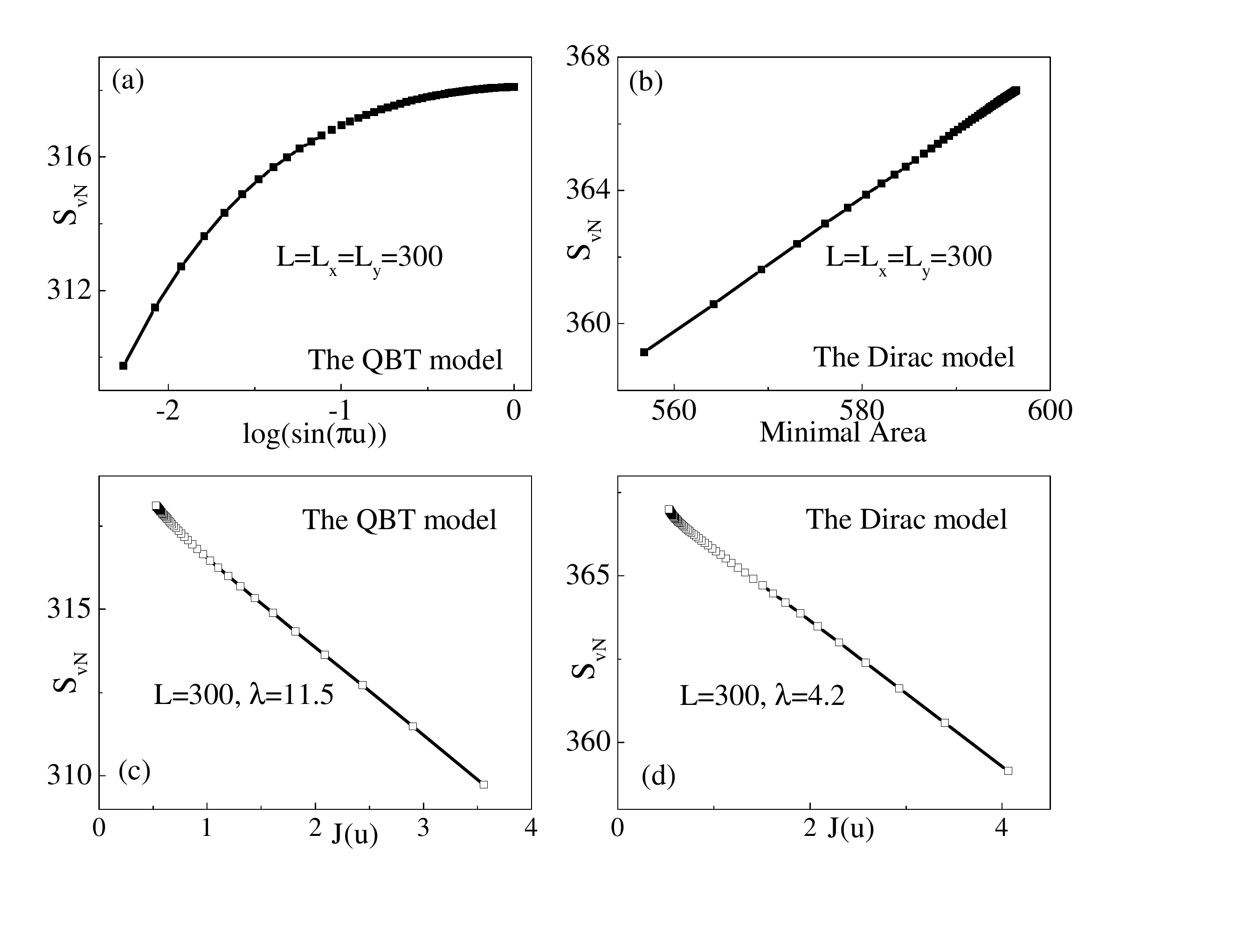}
\caption{(a) $S_{vN}$ for the QBT model in the function of $\log(\sin(\pi \y))$. (b) $S_{vN}$ for the Dirac model in the function of the minimal surface Eq.~\eqref{min_surface} for the AdS soliton geometry. (c) $S_{vN}$ for the QBT model in the function of $J(\y)$. (d) $S_{vN}$ for the Dirac model in the function of $J(\y)$. }
\label{fig:subleading_term}
\end{figure}

As shown in Fig.~\ref{fig:subleading_term} (c) and (d), when $L_y=L_x$ is fixed, for both the QBT and Dirac models, $S_{vN}$ is linearly related to $J(\y)$. For the Dirac model, $S_{vN}$ is also linearly proportional to $j(\y)$ as shown in Fig.~\ref{fig:subleading_term} (b). It is clearly seen from Fig.~\ref{fig:subleading_term} that the $J(\y)$ and $j(\y)$ fitting function works much better than the quasi-1D formula $\log(\sin(\y))$. The $\log(\sin(\pi\y))$ term which works well in the thin torus limit for the Dirac model is not linearly proportional to the numerical results in the two-dimensional limit (Fig.~\ref{fig:subleading_term} (a)).  For the $J(\y)$ function, there is an additional  tuning parameter $\lambda$. For the QBT model, $\lambda$ decreases when the R{\'e}nyi index $n$ increases, while for the Dirac model, $\lambda$ does not change when we increase the R{\'e}nyi index, and it is found to be equal to  $\lambda=4.2$. Currently, we do not have a physical understanding for the meaning of $\lambda$ for the fermionic models. In the case of the QLM,  $\lambda$ is the exponent of the two-point dimer correlation function\cite{Fradkin2013} and thus is independent of the R{\'e}nyi index $n>1$.\cite{Stephan2013b}

\begin{figure}[hbt]
\centering
\includegraphics[scale=.34]{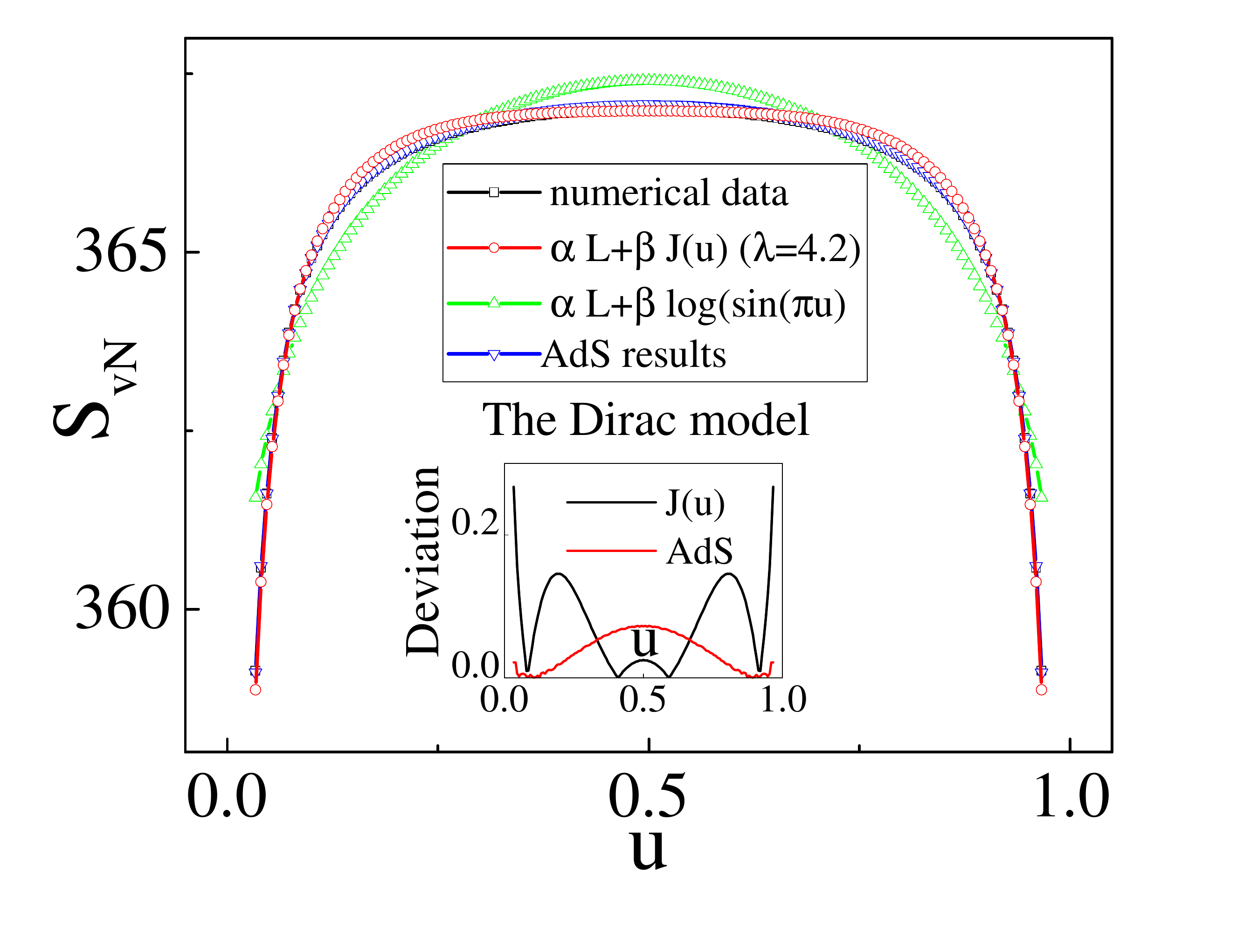}
\caption{$S_{vN}$ for the Dirac model as a function of $\y$ with $L=L_x=L_y=300$.  The red curve is the fitting function with the form $S_{vN}=\alpha L+\beta J(\y)$. The numerical data is in black curve. The blue curve is the holographic entropy. (The black and  blue curves are hard to see in the figure since they are almost overlapping with the blue curve) The inset is the absolute deviation for $S_{vN}=\alpha L+\beta J(\y)$ (black curve) and the holographic entropy (red curve) with the numerical data. In both cases, the deviation is less than $1\%$ for the whole region, but the holographic result appears to be the most accurate. The green curve is the fitting function with the form $S_{vN}= \alpha L + \beta \log(\sin(\pi \y))$.}
\label{fig:Dirac_model}
\end{figure}

\begin{figure}[hbt]
\centering
\includegraphics[scale=.34]{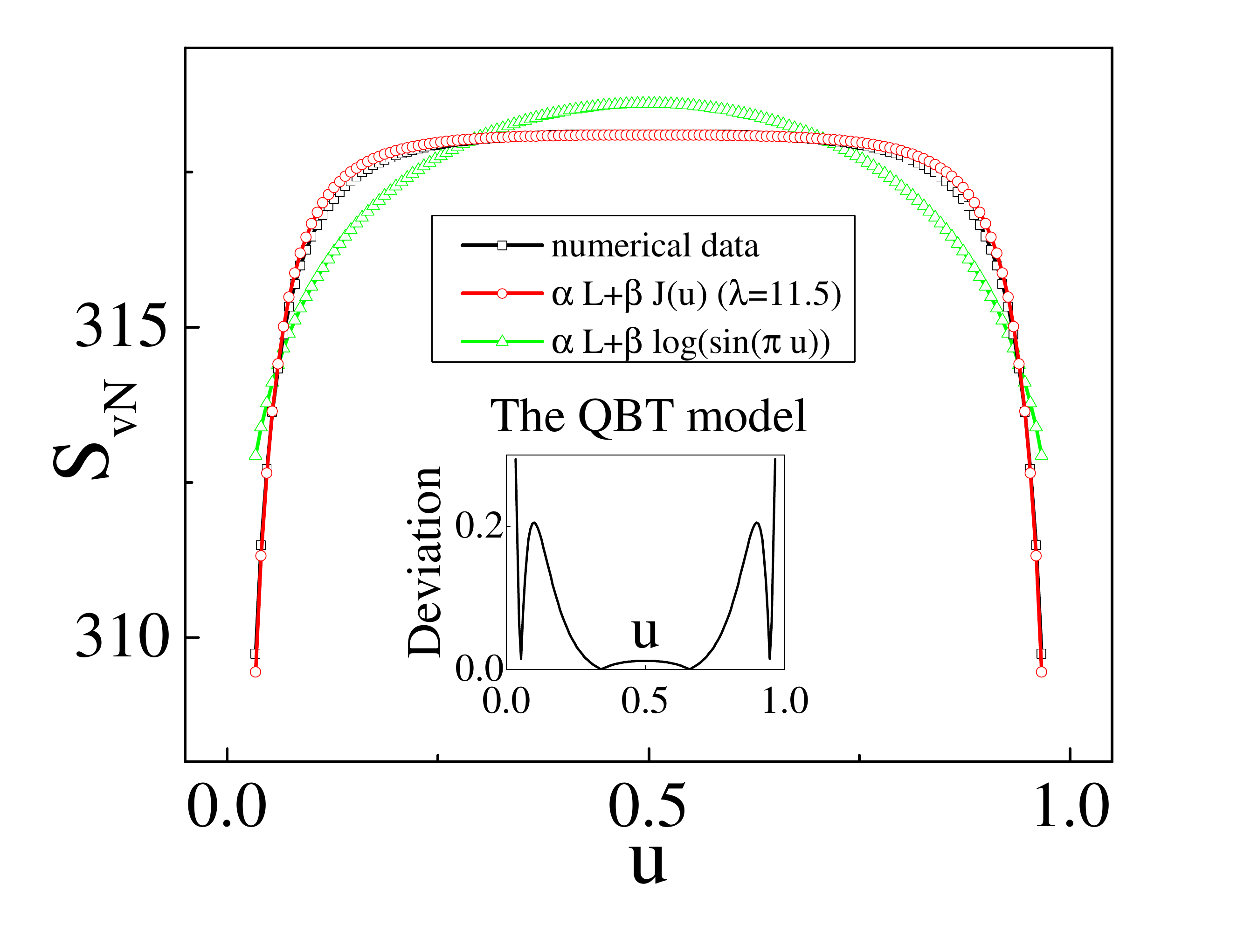}
\caption{$S_{vN}$ for the QBT model as a function of $\y$. The bipartition geometry is the same as the Dirac model. The inset is the absolute deviation for the fitting function $S_{vN}=\alpha L+\beta J(\y)$ with numerical data. The deviation is less than $1\%$ for the whole region of $\y$. }
\label{fig:QBT_model}
\end{figure}

The complete results are shown in Fig.~\ref{fig:Dirac_model} and Fig.~\ref{fig:QBT_model}, where the numerical calculations of $S_{vN}$ for both the QBT and Dirac models fit with $J(\y)$ for the whole range of the parameter $\y \in [0,1]$ within the numerical deviation $1\%$ (deviation shown  in the inset). Here we only study $S_{vN}$ because the other R{\'e}nyi entropies with different R{\'e}nyi indices show the similar behaviors. The only difference between the QBT and Dirac models is that for the QBT model, the fitting parameter $\lambda$ for $S_{vN}$ is $\lambda=11.5$ and for the Dirac model, $\lambda=4.2$, so that the curve for the QBT model is more flat around $\y=0.5$ compared with the Dirac model (Fig.~\ref{fig:Dirac_model} and Fig.~\ref{fig:QBT_model}). 

Furthermore, we notice that when $\y$ is small, $J(\y)$ is linearly proportional to $1/\y$. Similarly when $(L_x-L_A)\ll L_x$, $J(\y)$ is linearly proportional to $1/(1-\y)$. $J(\y)$ is symmetric around $\y=0.5$. To illustrate this, we expand $J(\y)$ for small $\y <<1$, to obtain
\begin{eqnarray}
J(\y)\approx\frac{\pi}{24}\frac{L_y}{L_x}\frac{1}{\y}+\log\left[\sqrt{\frac{2}{\lambda}}\frac{\theta_3(\lambda\tau)}{\eta(2\tau)}\right].
\end{eqnarray}
The leading term in $J(\y)$ is linearly proportional to $1/\y$ and independent of $\lambda$. This result is consistent with the numerical results and the holographic entropy derived before. 
Indeed as we mentioned previously the coefficient of the $1/\y$ term as $\y \rightarrow 0$ is intrinsic to the critical theory, 
and it is a rough analog of the central charge in $1+1$-dimensional CFT.
This statement can be further supported from the AdS/CFT calculation, where the coefficient only depends on the Newton constant $G_N$ (see Eq.~(\ref{HolographicEE})). 
For the lattice Dirac model (which has two species of massless Dirac fermions), the numerical data shows that the coefficient of the $1/\y$ is $-0.3006$, while for the QBT model, 
the coefficient is $-0.3735$. See Refs.
[\onlinecite{Ryu:2006ef,Myers:2012ed}] for related  studies of this quantity and how it flows under the RG. 
We use this coefficient to normalize the overall numerical coefficients $\beta$ of the different scaling functions we compare.

\section{Conclusion and comments}

The properties of a many-body state can be classified according to the scaling behavior of its EE. The critical system in $2+1$ dimensions is long-range entangled compared with the gapped system. Our results show that  this difference can be detected in the subleading term in the two-cylinder EE, which includes both the von Neumann entropy and R{\'e}nyi entropies. We calculate EEs of both the Dirac and quadratic band touching models numerically on the torus. We notice that the subleading term is linearly proportional to $1/\y$ when the ratio $\y$ is small. We speculate that the coefficient in front of $1/\y$ in the subleading term 
measures the number of the low-energy degrees of freedom of the system.  Further calculations in other models are necessary to pin down the physical meaning of this coefficient. For the whole region $0<\y<1$, we use the subleading term for the QLM and find it fit well with both the QBT model and the Dirac model within a small numeric deviation $<1\%$, even though these theories have different dynamical exponents, different DOS at low energies, and different behavior when the local four-fermion interactions are considered. We demonstrated that this similarity might come from the similar scaling behaviors between the two-point correlation functions at equal time for both the models. We also calculated the subleading term of the strongly-coupled models via the holographic AdS/CFT correspondence and find it consistent with the numerical results for the Dirac model. 

Based on our calculation on the fermionic critical models, holographic calculations and previous works on the bosonic critical models in $2+1$ dimensions, there is strong evidence that the
scaling form of the subleading term of EEs takes a robust form across a wide variety of $2+1$-dimensional critical systems on the torus geometry. It will be particularly interesting to test the holographic EE scaling we found here  against other critical theories such as the quantum Ising model in $2+1$ dimensions. 

\textit{Note}: Upon finishing this manuscript, we noticed a recent work by A. C. Potter which, among other topics, discusses the validity of the area law for EE in the QBT model based on the argument relating particle number fluctuations and EEs.\cite{Potter2014}

\begin{acknowledgments}
We thank Roger G. Melko for the discussion on the universal subleading terms and for pointing us to their work on the corner contributions in critical $2+1$-dimensional systems. We thank P. Fendley, S. Hartnoll, H.W. Kim, I. H. Kim, S.-S. Lee, J. Moore, A. Potter, S. Ryu, B. Swingle, S. Sachdev, N. Tubman, and H. Yao for discussions and comments. This work was supported in part by the National Science Foundation grants DMR-1064319 and DMR-1408713 (XC,GYC,EF)  at  the University of Illinois. 
\end{acknowledgments}

%


\begin{thebibliography}{73}%
\makeatletter
\providecommand \@ifxundefined [1]{%
 \@ifx{#1\undefined}
}%
\providecommand \@ifnum [1]{%
 \ifnum #1\expandafter \@firstoftwo
 \else \expandafter \@secondoftwo
 \fi
}%
\providecommand \@ifx [1]{%
 \ifx #1\expandafter \@firstoftwo
 \else \expandafter \@secondoftwo
 \fi
}%
\providecommand \natexlab [1]{#1}%
\providecommand \enquote  [1]{``#1''}%
\providecommand \bibnamefont  [1]{#1}%
\providecommand \bibfnamefont [1]{#1}%
\providecommand \citenamefont [1]{#1}%
\providecommand \href@noop [0]{\@secondoftwo}%
\providecommand \href [0]{\begingroup \@sanitize@url \@href}%
\providecommand \@href[1]{\@@startlink{#1}\@@href}%
\providecommand \@@href[1]{\endgroup#1\@@endlink}%
\providecommand \@sanitize@url [0]{\catcode `\\12\catcode `\$12\catcode
  `\&12\catcode `\#12\catcode `\^12\catcode `\_12\catcode `\%12\relax}%
\providecommand \@@startlink[1]{}%
\providecommand \@@endlink[0]{}%
\providecommand \url  [0]{\begingroup\@sanitize@url \@url }%
\providecommand \@url [1]{\endgroup\@href {#1}{\urlprefix }}%
\providecommand \urlprefix  [0]{URL }%
\providecommand \Eprint [0]{\href }%
\providecommand \doibase [0]{http://dx.doi.org/}%
\providecommand \selectlanguage [0]{\@gobble}%
\providecommand \bibinfo  [0]{\@secondoftwo}%
\providecommand \bibfield  [0]{\@secondoftwo}%
\providecommand \translation [1]{[#1]}%
\providecommand \BibitemOpen [0]{}%
\providecommand \bibitemStop [0]{}%
\providecommand \bibitemNoStop [0]{.\EOS\space}%
\providecommand \EOS [0]{\spacefactor3000\relax}%
\providecommand \BibitemShut  [1]{\csname bibitem#1\endcsname}%
\let\auto@bib@innerbib\@empty
\bibitem [{\citenamefont {Bombelli}\ \emph {et~al.}(1986)\citenamefont
  {Bombelli}, \citenamefont {Koul}, \citenamefont {Lee},\ and\ \citenamefont
  {Sorkin}}]{Bombelli-1986}%
  \BibitemOpen
  \bibfield  {author} {\bibinfo {author} {\bibfnamefont {L.}~\bibnamefont
  {Bombelli}}, \bibinfo {author} {\bibfnamefont {R.~K.}\ \bibnamefont {Koul}},
  \bibinfo {author} {\bibfnamefont {J.}~\bibnamefont {Lee}}, \ and\ \bibinfo
  {author} {\bibfnamefont {R.~D.}\ \bibnamefont {Sorkin}},\ }\href@noop {}
  {\bibfield  {journal} {\bibinfo  {journal} {Physical Review D}\ }\textbf
  {\bibinfo {volume} {34}},\ \bibinfo {pages} {373} (\bibinfo {year}
  {1986})}\BibitemShut {NoStop}%
\bibitem [{\citenamefont {Srednicki}(1993)}]{Srednicki-1993}%
  \BibitemOpen
  \bibfield  {author} {\bibinfo {author} {\bibfnamefont {M.}~\bibnamefont
  {Srednicki}},\ }\href@noop {} {\bibfield  {journal} {\bibinfo  {journal}
  {Physical Review Letters}\ }\textbf {\bibinfo {volume} {71}},\ \bibinfo
  {pages} {666} (\bibinfo {year} {1993})}\BibitemShut {NoStop}%
\bibitem [{\citenamefont {Wolf}\ \emph {et~al.}(2008)\citenamefont {Wolf},
  \citenamefont {Verstraete}, \citenamefont {Hastings},\ and\ \citenamefont
  {Cirac}}]{Wolf-2008}%
  \BibitemOpen
  \bibfield  {author} {\bibinfo {author} {\bibfnamefont {M.~M.}\ \bibnamefont
  {Wolf}}, \bibinfo {author} {\bibfnamefont {F.}~\bibnamefont {Verstraete}},
  \bibinfo {author} {\bibfnamefont {M.~B.}\ \bibnamefont {Hastings}}, \ and\
  \bibinfo {author} {\bibfnamefont {J.~I.}\ \bibnamefont {Cirac}},\ }\href@noop
  {} {\bibfield  {journal} {\bibinfo  {journal} {Physical Review Letters}\
  }\textbf {\bibinfo {volume} {100}},\ \bibinfo {pages} {070502} (\bibinfo
  {year} {2008})}\BibitemShut {NoStop}%
\bibitem [{\citenamefont {Ryu}\ and\ \citenamefont
  {Takayanagi}(2006{\natexlab{a}})}]{Ryu2006}%
  \BibitemOpen
  \bibfield  {author} {\bibinfo {author} {\bibfnamefont {S.}~\bibnamefont
  {Ryu}}\ and\ \bibinfo {author} {\bibfnamefont {T.}~\bibnamefont
  {Takayanagi}},\ }\href@noop {} {\bibfield  {journal} {\bibinfo  {journal}
  {Physical Review Letters}\ }\textbf {\bibinfo {volume} {96}},\ \bibinfo
  {pages} {181602} (\bibinfo {year} {2006}{\natexlab{a}})}\BibitemShut
  {NoStop}%
\bibitem [{\citenamefont {Casini}\ and\ \citenamefont
  {Huerta}(2009)}]{Casini-2009}%
  \BibitemOpen
  \bibfield  {author} {\bibinfo {author} {\bibfnamefont {H.}~\bibnamefont
  {Casini}}\ and\ \bibinfo {author} {\bibfnamefont {M.}~\bibnamefont
  {Huerta}},\ }\href@noop {} {\bibfield  {journal} {\bibinfo  {journal}
  {Journal of Physics A: Mathematical and Theoretical}\ }\textbf {\bibinfo
  {volume} {42}},\ \bibinfo {pages} {504007} (\bibinfo {year}
  {2009})}\BibitemShut {NoStop}%
\bibitem [{\citenamefont {Amico}\ \emph {et~al.}(2008)\citenamefont {Amico},
  \citenamefont {Fazio}, \citenamefont {Osterloh},\ and\ \citenamefont
  {Vedral}}]{Amico2008}%
  \BibitemOpen
  \bibfield  {author} {\bibinfo {author} {\bibfnamefont {L.}~\bibnamefont
  {Amico}}, \bibinfo {author} {\bibfnamefont {R.}~\bibnamefont {Fazio}},
  \bibinfo {author} {\bibfnamefont {A.}~\bibnamefont {Osterloh}}, \ and\
  \bibinfo {author} {\bibfnamefont {V.}~\bibnamefont {Vedral}},\ }\href@noop {}
  {\bibfield  {journal} {\bibinfo  {journal} {Reviews of Modern Physics}\
  }\textbf {\bibinfo {volume} {80}},\ \bibinfo {pages} {517} (\bibinfo {year}
  {2008})}\BibitemShut {NoStop}%
\bibitem [{\citenamefont {Eisert}\ \emph {et~al.}(2010)\citenamefont {Eisert},
  \citenamefont {Cramer},\ and\ \citenamefont {Plenio}}]{Eisert-2010}%
  \BibitemOpen
  \bibfield  {author} {\bibinfo {author} {\bibfnamefont {J.}~\bibnamefont
  {Eisert}}, \bibinfo {author} {\bibfnamefont {M.}~\bibnamefont {Cramer}}, \
  and\ \bibinfo {author} {\bibfnamefont {M.~B.}\ \bibnamefont {Plenio}},\
  }\href@noop {} {\bibfield  {journal} {\bibinfo  {journal} {Reviews of Modern
  Physics}\ }\textbf {\bibinfo {volume} {82}},\ \bibinfo {pages} {277}
  (\bibinfo {year} {2010})}\BibitemShut {NoStop}%
\bibitem [{\citenamefont {Fradkin}(2013)}]{Fradkin2013}%
  \BibitemOpen
  \bibfield  {author} {\bibinfo {author} {\bibfnamefont {E.}~\bibnamefont
  {Fradkin}},\ }\href@noop {} {\emph {\bibinfo {title} {Field Theories of
  Condensed Matter Systems}}}\ (\bibinfo  {publisher} {Cambridge University
  Press},\ \bibinfo {address} {Cambridge, UK},\ \bibinfo {year}
  {2013})\BibitemShut {NoStop}%
\bibitem [{\citenamefont {Callan}\ and\ \citenamefont
  {Wilczek}(1994)}]{Callan1994}%
  \BibitemOpen
  \bibfield  {author} {\bibinfo {author} {\bibfnamefont {C.~G.}\ \bibnamefont
  {Callan}}\ and\ \bibinfo {author} {\bibfnamefont {F.}~\bibnamefont
  {Wilczek}},\ }\href@noop {} {\bibfield  {journal} {\bibinfo  {journal}
  {Physics Letters B}\ }\textbf {\bibinfo {volume} {333}},\ \bibinfo {pages}
  {55} (\bibinfo {year} {1994})}\BibitemShut {NoStop}%
\bibitem [{\citenamefont {Holzhey}\ \emph {et~al.}(1994)\citenamefont
  {Holzhey}, \citenamefont {Larsen},\ and\ \citenamefont
  {Wilczek}}]{Holzhey1994}%
  \BibitemOpen
  \bibfield  {author} {\bibinfo {author} {\bibfnamefont {C.}~\bibnamefont
  {Holzhey}}, \bibinfo {author} {\bibfnamefont {F.}~\bibnamefont {Larsen}}, \
  and\ \bibinfo {author} {\bibfnamefont {F.}~\bibnamefont {Wilczek}},\
  }\href@noop {} {\bibfield  {journal} {\bibinfo  {journal} {Nuclear Physics
  B}\ }\textbf {\bibinfo {volume} {424}},\ \bibinfo {pages} {443} (\bibinfo
  {year} {1994})}\BibitemShut {NoStop}%
\bibitem [{\citenamefont {Calabrese}\ and\ \citenamefont
  {Cardy}(2004)}]{Calabrese2004}%
  \BibitemOpen
  \bibfield  {author} {\bibinfo {author} {\bibfnamefont {P.}~\bibnamefont
  {Calabrese}}\ and\ \bibinfo {author} {\bibfnamefont {J.}~\bibnamefont
  {Cardy}},\ }\href@noop {} {\bibfield  {journal} {\bibinfo  {journal} {Journal
  of Statistical Mechanics}\ }\textbf {\bibinfo {volume} {2004}},\ \bibinfo
  {pages} {P06002} (\bibinfo {year} {2004})}\BibitemShut {NoStop}%
\bibitem [{\citenamefont {Calabrese}\ and\ \citenamefont
  {Cardy}(2009)}]{Calabrese-2009b}%
  \BibitemOpen
  \bibfield  {author} {\bibinfo {author} {\bibfnamefont {P.}~\bibnamefont
  {Calabrese}}\ and\ \bibinfo {author} {\bibfnamefont {J.}~\bibnamefont
  {Cardy}},\ }\href@noop {} {\bibfield  {journal} {\bibinfo  {journal} {Journal
  of Physics A: Theoretical and Mathematical}\ }\textbf {\bibinfo {volume}
  {42}},\ \bibinfo {pages} {504005} (\bibinfo {year} {2009})}\BibitemShut
  {NoStop}%
\bibitem [{\citenamefont {Caraglio}\ and\ \citenamefont
  {Gliozzi}(2008)}]{Caraglio-2008}%
  \BibitemOpen
  \bibfield  {author} {\bibinfo {author} {\bibfnamefont {M.}~\bibnamefont
  {Caraglio}}\ and\ \bibinfo {author} {\bibfnamefont {F.}~\bibnamefont
  {Gliozzi}},\ }\href@noop {} {\bibfield  {journal} {\bibinfo  {journal}
  {Journal of High Energy Physics}\ }\textbf {\bibinfo {volume} {2008}},\
  \bibinfo {pages} {076} (\bibinfo {year} {2008})}\BibitemShut {NoStop}%
\bibitem [{\citenamefont {Calabrese}\ \emph {et~al.}(2009)\citenamefont
  {Calabrese}, \citenamefont {Cardy},\ and\ \citenamefont
  {Tonni}}]{Calabrese-2009}%
  \BibitemOpen
  \bibfield  {author} {\bibinfo {author} {\bibfnamefont {P.}~\bibnamefont
  {Calabrese}}, \bibinfo {author} {\bibfnamefont {J.}~\bibnamefont {Cardy}}, \
  and\ \bibinfo {author} {\bibfnamefont {E.}~\bibnamefont {Tonni}},\
  }\href@noop {} {\bibfield  {journal} {\bibinfo  {journal} {Journal of
  Statistical Mechanics}\ }\textbf {\bibinfo {volume} {2009}},\ \bibinfo
  {pages} {P11001} (\bibinfo {year} {2009})}\BibitemShut {NoStop}%
\bibitem [{\citenamefont {Calabrese}\ \emph
  {et~al.}(2012{\natexlab{a}})\citenamefont {Calabrese}, \citenamefont
  {Cardy},\ and\ \citenamefont {Tonni}}]{Calabrese2012}%
  \BibitemOpen
  \bibfield  {author} {\bibinfo {author} {\bibfnamefont {P.}~\bibnamefont
  {Calabrese}}, \bibinfo {author} {\bibfnamefont {J.}~\bibnamefont {Cardy}}, \
  and\ \bibinfo {author} {\bibfnamefont {E.}~\bibnamefont {Tonni}},\
  }\href@noop {} {\bibfield  {journal} {\bibinfo  {journal} {Physical Review
  Letters}\ }\textbf {\bibinfo {volume} {109}},\ \bibinfo {pages} {130502}
  (\bibinfo {year} {2012}{\natexlab{a}})}\BibitemShut {NoStop}%
\bibitem [{\citenamefont {Cardy}(2013)}]{Cardy-2013}%
  \BibitemOpen
  \bibfield  {author} {\bibinfo {author} {\bibfnamefont {J.}~\bibnamefont
  {Cardy}},\ }\href@noop {} {\bibfield  {journal} {\bibinfo  {journal} {Journal
  of Physics A: Mathematical and Theoretical}\ }\textbf {\bibinfo {volume}
  {46}},\ \bibinfo {pages} {285402} (\bibinfo {year} {2013})}\BibitemShut
  {NoStop}%
\bibitem [{\citenamefont {Kitaev}\ and\ \citenamefont
  {Preskill}(2006)}]{Kitaev2006}%
  \BibitemOpen
  \bibfield  {author} {\bibinfo {author} {\bibfnamefont {A.}~\bibnamefont
  {Kitaev}}\ and\ \bibinfo {author} {\bibfnamefont {J.}~\bibnamefont
  {Preskill}},\ }\href@noop {} {\bibfield  {journal} {\bibinfo  {journal}
  {Physical Review Letters}\ }\textbf {\bibinfo {volume} {96}},\ \bibinfo
  {pages} {110404} (\bibinfo {year} {2006})}\BibitemShut {NoStop}%
\bibitem [{\citenamefont {Levin}\ and\ \citenamefont {Wen}(2006)}]{Levin2006}%
  \BibitemOpen
  \bibfield  {author} {\bibinfo {author} {\bibfnamefont {M.}~\bibnamefont
  {Levin}}\ and\ \bibinfo {author} {\bibfnamefont {X.-G.}\ \bibnamefont
  {Wen}},\ }\href@noop {} {\bibfield  {journal} {\bibinfo  {journal} {Physical
  Review Letters}\ }\textbf {\bibinfo {volume} {96}},\ \bibinfo {pages}
  {110405} (\bibinfo {year} {2006})}\BibitemShut {NoStop}%
\bibitem [{\citenamefont {Dong}\ \emph {et~al.}(2008)\citenamefont {Dong},
  \citenamefont {Fradkin}, \citenamefont {Leigh},\ and\ \citenamefont
  {Nowling}}]{Dong-2008}%
  \BibitemOpen
  \bibfield  {author} {\bibinfo {author} {\bibfnamefont {S.}~\bibnamefont
  {Dong}}, \bibinfo {author} {\bibfnamefont {E.}~\bibnamefont {Fradkin}},
  \bibinfo {author} {\bibfnamefont {R.~G.}\ \bibnamefont {Leigh}}, \ and\
  \bibinfo {author} {\bibfnamefont {S.}~\bibnamefont {Nowling}},\ }\href@noop
  {} {\bibfield  {journal} {\bibinfo  {journal} {Journal of High Energy
  Physics}\ }\textbf {\bibinfo {volume} {05}},\ \bibinfo {pages} {016}
  (\bibinfo {year} {2008})}\BibitemShut {NoStop}%
\bibitem [{\citenamefont {Zozulya}\ \emph {et~al.}(2009)\citenamefont
  {Zozulya}, \citenamefont {Haque},\ and\ \citenamefont
  {Regnault}}]{Zozulya-2009}%
  \BibitemOpen
  \bibfield  {author} {\bibinfo {author} {\bibfnamefont {O.~S.}\ \bibnamefont
  {Zozulya}}, \bibinfo {author} {\bibfnamefont {M.}~\bibnamefont {Haque}}, \
  and\ \bibinfo {author} {\bibfnamefont {N.}~\bibnamefont {Regnault}},\
  }\href@noop {} {\bibfield  {journal} {\bibinfo  {journal} {Physical Review
  B}\ }\textbf {\bibinfo {volume} {79}},\ \bibinfo {pages} {045409} (\bibinfo
  {year} {2009})}\BibitemShut {NoStop}%
\bibitem [{\citenamefont {Zozulya}\ \emph {et~al.}(2007)\citenamefont
  {Zozulya}, \citenamefont {Haque}, \citenamefont {Schoutens},\ and\
  \citenamefont {Rezayi}}]{Zozulya-2007}%
  \BibitemOpen
  \bibfield  {author} {\bibinfo {author} {\bibfnamefont {O.~S.}\ \bibnamefont
  {Zozulya}}, \bibinfo {author} {\bibfnamefont {M.}~\bibnamefont {Haque}},
  \bibinfo {author} {\bibfnamefont {K.}~\bibnamefont {Schoutens}}, \ and\
  \bibinfo {author} {\bibfnamefont {E.~H.}\ \bibnamefont {Rezayi}},\
  }\href@noop {} {\bibfield  {journal} {\bibinfo  {journal} {Physical Review
  B}\ }\textbf {\bibinfo {volume} {76}},\ \bibinfo {pages} {125310} (\bibinfo
  {year} {2007})}\BibitemShut {NoStop}%
\bibitem [{\citenamefont {Li}\ and\ \citenamefont
  {Haldane}(2008)}]{Haldane-2008}%
  \BibitemOpen
  \bibfield  {author} {\bibinfo {author} {\bibfnamefont {H.}~\bibnamefont
  {Li}}\ and\ \bibinfo {author} {\bibfnamefont {F.}~\bibnamefont {Haldane}},\
  }\href@noop {} {\bibfield  {journal} {\bibinfo  {journal} {Physical Review
  Letters}\ }\textbf {\bibinfo {volume} {101}},\ \bibinfo {pages} {010504}
  (\bibinfo {year} {2008})}\BibitemShut {NoStop}%
\bibitem [{\citenamefont {Furukawa}\ and\ \citenamefont
  {Misguich}(2007)}]{Furukawa-2007}%
  \BibitemOpen
  \bibfield  {author} {\bibinfo {author} {\bibfnamefont {S.}~\bibnamefont
  {Furukawa}}\ and\ \bibinfo {author} {\bibfnamefont {G.}~\bibnamefont
  {Misguich}},\ }\href@noop {} {\bibfield  {journal} {\bibinfo  {journal}
  {Physical Review B}\ }\textbf {\bibinfo {volume} {75}},\ \bibinfo {pages}
  {214407} (\bibinfo {year} {2007})}\BibitemShut {NoStop}%
\bibitem [{\citenamefont {Papanikolaou}\ \emph {et~al.}(2007)\citenamefont
  {Papanikolaou}, \citenamefont {Raman},\ and\ \citenamefont
  {Fradkin}}]{Papanikolaou-2007}%
  \BibitemOpen
  \bibfield  {author} {\bibinfo {author} {\bibfnamefont {S.}~\bibnamefont
  {Papanikolaou}}, \bibinfo {author} {\bibfnamefont {K.~S.}\ \bibnamefont
  {Raman}}, \ and\ \bibinfo {author} {\bibfnamefont {E.}~\bibnamefont
  {Fradkin}},\ }\href@noop {} {\bibfield  {journal} {\bibinfo  {journal}
  {Physical Review B}\ }\textbf {\bibinfo {volume} {76}},\ \bibinfo {pages}
  {224421} (\bibinfo {year} {2007})}\BibitemShut {NoStop}%
\bibitem [{\citenamefont {St\'ephan}\ \emph {et~al.}(2009)\citenamefont
  {St\'ephan}, \citenamefont {Furukawa}, \citenamefont {Misguich},\ and\
  \citenamefont {Pasquier}}]{Stephan-2009}%
  \BibitemOpen
  \bibfield  {author} {\bibinfo {author} {\bibfnamefont {J.~M.}\ \bibnamefont
  {St\'ephan}}, \bibinfo {author} {\bibfnamefont {S.}~\bibnamefont {Furukawa}},
  \bibinfo {author} {\bibfnamefont {G.}~\bibnamefont {Misguich}}, \ and\
  \bibinfo {author} {\bibfnamefont {V.}~\bibnamefont {Pasquier}},\ }\href@noop
  {} {\bibfield  {journal} {\bibinfo  {journal} {Physical Review B}\ }\textbf
  {\bibinfo {volume} {80}},\ \bibinfo {pages} {184421} (\bibinfo {year}
  {2009})}\BibitemShut {NoStop}%
\bibitem [{\citenamefont {Hamma}\ \emph
  {et~al.}(2005{\natexlab{a}})\citenamefont {Hamma}, \citenamefont
  {Ionicioiu},\ and\ \citenamefont {Zanardi}}]{Hamma-2005}%
  \BibitemOpen
  \bibfield  {author} {\bibinfo {author} {\bibfnamefont {A.}~\bibnamefont
  {Hamma}}, \bibinfo {author} {\bibfnamefont {R.}~\bibnamefont {Ionicioiu}}, \
  and\ \bibinfo {author} {\bibfnamefont {P.}~\bibnamefont {Zanardi}},\
  }\href@noop {} {\bibfield  {journal} {\bibinfo  {journal} {Physics Letters
  A}\ }\textbf {\bibinfo {volume} {337}},\ \bibinfo {pages} {22} (\bibinfo
  {year} {2005}{\natexlab{a}})}\BibitemShut {NoStop}%
\bibitem [{\citenamefont {Hamma}\ \emph
  {et~al.}(2005{\natexlab{b}})\citenamefont {Hamma}, \citenamefont
  {Ionicioiu},\ and\ \citenamefont {Zanardi}}]{Hamma-2005b}%
  \BibitemOpen
  \bibfield  {author} {\bibinfo {author} {\bibfnamefont {A.}~\bibnamefont
  {Hamma}}, \bibinfo {author} {\bibfnamefont {R.}~\bibnamefont {Ionicioiu}}, \
  and\ \bibinfo {author} {\bibfnamefont {P.}~\bibnamefont {Zanardi}},\
  }\href@noop {} {\bibfield  {journal} {\bibinfo  {journal} {Physical Review
  A}\ }\textbf {\bibinfo {volume} {71}},\ \bibinfo {pages} {022315} (\bibinfo
  {year} {2005}{\natexlab{b}})}\BibitemShut {NoStop}%
\bibitem [{\citenamefont {Castelnovo}\ and\ \citenamefont
  {Chamon}(2008)}]{Castelnovo-2008}%
  \BibitemOpen
  \bibfield  {author} {\bibinfo {author} {\bibfnamefont {C.}~\bibnamefont
  {Castelnovo}}\ and\ \bibinfo {author} {\bibfnamefont {C.}~\bibnamefont
  {Chamon}},\ }\href@noop {} {\bibfield  {journal} {\bibinfo  {journal}
  {Physical Review B}\ }\textbf {\bibinfo {volume} {77}},\ \bibinfo {pages}
  {054433} (\bibinfo {year} {2008})}\BibitemShut {NoStop}%
\bibitem [{\citenamefont {Zhang}\ \emph {et~al.}(2012)\citenamefont {Zhang},
  \citenamefont {Grover}, \citenamefont {Turner}, \citenamefont {Oshikawa},\
  and\ \citenamefont {Vishwanath}}]{Yi-2012}%
  \BibitemOpen
  \bibfield  {author} {\bibinfo {author} {\bibfnamefont {Y.}~\bibnamefont
  {Zhang}}, \bibinfo {author} {\bibfnamefont {T.}~\bibnamefont {Grover}},
  \bibinfo {author} {\bibfnamefont {A.}~\bibnamefont {Turner}}, \bibinfo
  {author} {\bibfnamefont {M.}~\bibnamefont {Oshikawa}}, \ and\ \bibinfo
  {author} {\bibfnamefont {A.}~\bibnamefont {Vishwanath}},\ }\href@noop {}
  {\bibfield  {journal} {\bibinfo  {journal} {Physical Review B}\ }\textbf
  {\bibinfo {volume} {85}},\ \bibinfo {pages} {235151} (\bibinfo {year}
  {2012})}\BibitemShut {NoStop}%
\bibitem [{\citenamefont {Fradkin}\ and\ \citenamefont
  {Moore}(2006)}]{Fradkin2006}%
  \BibitemOpen
  \bibfield  {author} {\bibinfo {author} {\bibfnamefont {E.}~\bibnamefont
  {Fradkin}}\ and\ \bibinfo {author} {\bibfnamefont {J.}~\bibnamefont
  {Moore}},\ }\href@noop {} {\bibfield  {journal} {\bibinfo  {journal}
  {Physical Review Letters}\ }\textbf {\bibinfo {volume} {97}},\ \bibinfo
  {pages} {050404} (\bibinfo {year} {2006})}\BibitemShut {NoStop}%
\bibitem [{\citenamefont {Casini}\ and\ \citenamefont
  {Huerta}(2007)}]{Casini-2007}%
  \BibitemOpen
  \bibfield  {author} {\bibinfo {author} {\bibfnamefont {H.}~\bibnamefont
  {Casini}}\ and\ \bibinfo {author} {\bibfnamefont {M.}~\bibnamefont
  {Huerta}},\ }\href@noop {} {\bibfield  {journal} {\bibinfo  {journal}
  {Nuclear Physics B}\ }\textbf {\bibinfo {volume} {764}},\ \bibinfo {pages}
  {183} (\bibinfo {year} {2007})}\BibitemShut {NoStop}%
\bibitem [{\citenamefont {Hsu}\ \emph {et~al.}(2009)\citenamefont {Hsu},
  \citenamefont {Mulligan}, \citenamefont {Fradkin},\ and\ \citenamefont
  {Kim}}]{Hsu2009}%
  \BibitemOpen
  \bibfield  {author} {\bibinfo {author} {\bibfnamefont {B.}~\bibnamefont
  {Hsu}}, \bibinfo {author} {\bibfnamefont {M.}~\bibnamefont {Mulligan}},
  \bibinfo {author} {\bibfnamefont {E.}~\bibnamefont {Fradkin}}, \ and\
  \bibinfo {author} {\bibfnamefont {E.-A.}\ \bibnamefont {Kim}},\ }\href@noop
  {} {\bibfield  {journal} {\bibinfo  {journal} {Physical Review B}\ }\textbf
  {\bibinfo {volume} {79}},\ \bibinfo {pages} {115421} (\bibinfo {year}
  {2009})}\BibitemShut {NoStop}%
\bibitem [{\citenamefont {Hsu}\ and\ \citenamefont {Fradkin}(2010)}]{Hsu2010}%
  \BibitemOpen
  \bibfield  {author} {\bibinfo {author} {\bibfnamefont {B.}~\bibnamefont
  {Hsu}}\ and\ \bibinfo {author} {\bibfnamefont {E.}~\bibnamefont {Fradkin}},\
  }\href@noop {} {\bibfield  {journal} {\bibinfo  {journal} {Journal of
  Statistical Mechanics}\ }\textbf {\bibinfo {volume} {2010}},\ \bibinfo
  {pages} {P09004} (\bibinfo {year} {2010})}\BibitemShut {NoStop}%
\bibitem [{\citenamefont {St{\'e}phan}\ \emph {et~al.}(2011)\citenamefont
  {St{\'e}phan}, \citenamefont {Misguich},\ and\ \citenamefont
  {Pasquier}}]{Stephan2011}%
  \BibitemOpen
  \bibfield  {author} {\bibinfo {author} {\bibfnamefont {J.-M.}\ \bibnamefont
  {St{\'e}phan}}, \bibinfo {author} {\bibfnamefont {G.}~\bibnamefont
  {Misguich}}, \ and\ \bibinfo {author} {\bibfnamefont {V.}~\bibnamefont
  {Pasquier}},\ }\href@noop {} {\bibfield  {journal} {\bibinfo  {journal}
  {Physical Review B}\ }\textbf {\bibinfo {volume} {84}},\ \bibinfo {pages}
  {195128} (\bibinfo {year} {2011})}\BibitemShut {NoStop}%
\bibitem [{\citenamefont {St{\'e}phan}\ \emph {et~al.}(2012)\citenamefont
  {St{\'e}phan}, \citenamefont {Misguich},\ and\ \citenamefont
  {Pasquier}}]{Stephan-2012}%
  \BibitemOpen
  \bibfield  {author} {\bibinfo {author} {\bibfnamefont {J.~M.}\ \bibnamefont
  {St{\'e}phan}}, \bibinfo {author} {\bibfnamefont {G.}~\bibnamefont
  {Misguich}}, \ and\ \bibinfo {author} {\bibfnamefont {V.}~\bibnamefont
  {Pasquier}},\ }\href@noop {} {\bibfield  {journal} {\bibinfo  {journal}
  {Journal of Statistical Mechanics}\ }\textbf {\bibinfo {volume} {2012}},\
  \bibinfo {pages} {P02003} (\bibinfo {year} {2012})}\BibitemShut {NoStop}%
\bibitem [{\citenamefont {Ardonne}\ \emph {et~al.}(2004)\citenamefont
  {Ardonne}, \citenamefont {Fendley},\ and\ \citenamefont
  {Fradkin}}]{Ardonne2004}%
  \BibitemOpen
  \bibfield  {author} {\bibinfo {author} {\bibfnamefont {E.}~\bibnamefont
  {Ardonne}}, \bibinfo {author} {\bibfnamefont {P.}~\bibnamefont {Fendley}}, \
  and\ \bibinfo {author} {\bibfnamefont {E.}~\bibnamefont {Fradkin}},\
  }\href@noop {} {\bibfield  {journal} {\bibinfo  {journal} {Annals of
  Physics}\ }\textbf {\bibinfo {volume} {310}},\ \bibinfo {pages} {493}
  (\bibinfo {year} {2004})}\BibitemShut {NoStop}%
\bibitem [{\citenamefont {Fradkin}\ \emph {et~al.}(2004)\citenamefont
  {Fradkin}, \citenamefont {Huse}, \citenamefont {Moessner}, \citenamefont
  {Oganesyan},\ and\ \citenamefont {Sondhi}}]{Fradkin-2004}%
  \BibitemOpen
  \bibfield  {author} {\bibinfo {author} {\bibfnamefont {E.}~\bibnamefont
  {Fradkin}}, \bibinfo {author} {\bibfnamefont {D.}~\bibnamefont {Huse}},
  \bibinfo {author} {\bibfnamefont {R.}~\bibnamefont {Moessner}}, \bibinfo
  {author} {\bibfnamefont {V.}~\bibnamefont {Oganesyan}}, \ and\ \bibinfo
  {author} {\bibfnamefont {S.~L.}\ \bibnamefont {Sondhi}},\ }\href@noop {}
  {\bibfield  {journal} {\bibinfo  {journal} {Physical Review B}\ }\textbf
  {\bibinfo {volume} {69}},\ \bibinfo {pages} {224415} (\bibinfo {year}
  {2004})}\BibitemShut {NoStop}%
\bibitem [{\citenamefont {St{\'e}phan}\ \emph
  {et~al.}(2013{\natexlab{a}})\citenamefont {St{\'e}phan}, \citenamefont {Ju},
  \citenamefont {Fendley},\ and\ \citenamefont {Melko}}]{Stephan2013b}%
  \BibitemOpen
  \bibfield  {author} {\bibinfo {author} {\bibfnamefont {J.-M.}\ \bibnamefont
  {St{\'e}phan}}, \bibinfo {author} {\bibfnamefont {H.}~\bibnamefont {Ju}},
  \bibinfo {author} {\bibfnamefont {P.}~\bibnamefont {Fendley}}, \ and\
  \bibinfo {author} {\bibfnamefont {R.~G.}\ \bibnamefont {Melko}},\ }\href@noop
  {} {\bibfield  {journal} {\bibinfo  {journal} {New Journal of Physics}\
  }\textbf {\bibinfo {volume} {15}},\ \bibinfo {pages} {015004} (\bibinfo
  {year} {2013}{\natexlab{a}})}\BibitemShut {NoStop}%
\bibitem [{\citenamefont {Casini}\ and\ \citenamefont
  {Huerta}(2010)}]{Casini-2010}%
  \BibitemOpen
  \bibfield  {author} {\bibinfo {author} {\bibfnamefont {H.}~\bibnamefont
  {Casini}}\ and\ \bibinfo {author} {\bibfnamefont {M.}~\bibnamefont
  {Huerta}},\ }\href@noop {} {\bibfield  {journal} {\bibinfo  {journal}
  {Physics Letters B}\ }\textbf {\bibinfo {volume} {694}},\ \bibinfo {pages}
  {167} (\bibinfo {year} {2010})}\BibitemShut {NoStop}%
\bibitem [{\citenamefont {Klebanov}\ \emph {et~al.}(2011)\citenamefont
  {Klebanov}, \citenamefont {Pufu},\ and\ \citenamefont
  {Safdi}}]{Klebanov2011}%
  \BibitemOpen
  \bibfield  {author} {\bibinfo {author} {\bibfnamefont {I.~R.}\ \bibnamefont
  {Klebanov}}, \bibinfo {author} {\bibfnamefont {S.~S.}\ \bibnamefont {Pufu}},
  \ and\ \bibinfo {author} {\bibfnamefont {B.~R.}\ \bibnamefont {Safdi}},\
  }\href@noop {} {\bibfield  {journal} {\bibinfo  {journal} {Journal of High
  Energy Physics}\ }\textbf {\bibinfo {volume} {2011}},\ \bibinfo {pages} {38}
  (\bibinfo {year} {2011})}\BibitemShut {NoStop}%
\bibitem [{\citenamefont {Casini}\ \emph {et~al.}(2011)\citenamefont {Casini},
  \citenamefont {Huerta},\ and\ \citenamefont {Myers}}]{Casini-2011}%
  \BibitemOpen
  \bibfield  {author} {\bibinfo {author} {\bibfnamefont {H.}~\bibnamefont
  {Casini}}, \bibinfo {author} {\bibfnamefont {M.}~\bibnamefont {Huerta}}, \
  and\ \bibinfo {author} {\bibfnamefont {R.~C.}\ \bibnamefont {Myers}},\
  }\href@noop {} {\bibfield  {journal} {\bibinfo  {journal} {Journal of High
  Energy Physics}\ }\textbf {\bibinfo {volume} {2011}},\ \bibinfo {pages} {036}
  (\bibinfo {year} {2011})}\BibitemShut {NoStop}%
\bibitem [{\citenamefont {Casini}\ and\ \citenamefont
  {Huerta}(2012)}]{Casini-2012ei}%
  \BibitemOpen
  \bibfield  {author} {\bibinfo {author} {\bibfnamefont {H.}~\bibnamefont
  {Casini}}\ and\ \bibinfo {author} {\bibfnamefont {M.}~\bibnamefont
  {Huerta}},\ }\href@noop {} {\bibfield  {journal} {\bibinfo  {journal}
  {Physical Review D}\ }\textbf {\bibinfo {volume} {85}},\ \bibinfo {pages}
  {125016} (\bibinfo {year} {2012})}\BibitemShut {NoStop}%
\bibitem [{\citenamefont {Dowker}(2010)}]{Dowker-2010}%
  \BibitemOpen
  \bibfield  {author} {\bibinfo {author} {\bibfnamefont {J.~S.}\ \bibnamefont
  {Dowker}},\ }\href@noop {} {\enquote {\bibinfo {title} {{Entanglement entropy
  for odd spheres}},}\ } (\bibinfo {year} {2010}),\ \Eprint
  {http://arxiv.org/abs/arXiv:1012.1548} {arXiv:1012.1548} \BibitemShut
  {NoStop}%
\bibitem [{\citenamefont {Metlitski}\ \emph {et~al.}(2009)\citenamefont
  {Metlitski}, \citenamefont {Fuertes},\ and\ \citenamefont
  {Sachdev}}]{Metlitski}%
  \BibitemOpen
  \bibfield  {author} {\bibinfo {author} {\bibfnamefont {M.~A.}\ \bibnamefont
  {Metlitski}}, \bibinfo {author} {\bibfnamefont {C.~A.}\ \bibnamefont
  {Fuertes}}, \ and\ \bibinfo {author} {\bibfnamefont {S.}~\bibnamefont
  {Sachdev}},\ }\href@noop {} {\bibfield  {journal} {\bibinfo  {journal}
  {Physical Review B}\ }\textbf {\bibinfo {volume} {80}},\ \bibinfo {pages}
  {115122} (\bibinfo {year} {2009})}\BibitemShut {NoStop}%
\bibitem [{\citenamefont {Zaletel}\ \emph {et~al.}(2011)\citenamefont
  {Zaletel}, \citenamefont {Bardarson},\ and\ \citenamefont
  {Moore}}]{zalatel-2011}%
  \BibitemOpen
  \bibfield  {author} {\bibinfo {author} {\bibfnamefont {M.~P.}\ \bibnamefont
  {Zaletel}}, \bibinfo {author} {\bibfnamefont {J.~H.}\ \bibnamefont
  {Bardarson}}, \ and\ \bibinfo {author} {\bibfnamefont {J.~E.}\ \bibnamefont
  {Moore}},\ }\href@noop {} {\bibfield  {journal} {\bibinfo  {journal}
  {Physical Review Letters}\ }\textbf {\bibinfo {volume} {107}},\ \bibinfo
  {pages} {020402} (\bibinfo {year} {2011})}\BibitemShut {NoStop}%
\bibitem [{\citenamefont {Kallin}\ \emph {et~al.}(2014)\citenamefont {Kallin},
  \citenamefont {Stoudenmire}, \citenamefont {Fendley}, \citenamefont {Singh},\
  and\ \citenamefont {Melko}}]{Kallin-2014}%
  \BibitemOpen
  \bibfield  {author} {\bibinfo {author} {\bibfnamefont {A.~B.}\ \bibnamefont
  {Kallin}}, \bibinfo {author} {\bibfnamefont {E.~M.}\ \bibnamefont
  {Stoudenmire}}, \bibinfo {author} {\bibfnamefont {P.}~\bibnamefont
  {Fendley}}, \bibinfo {author} {\bibfnamefont {R.~R.~P.}\ \bibnamefont
  {Singh}}, \ and\ \bibinfo {author} {\bibfnamefont {R.~G.}\ \bibnamefont
  {Melko}},\ }\href@noop {} {\bibfield  {journal} {\bibinfo  {journal} {Journal
  of Statistical Mechanics}\ }\textbf {\bibinfo {volume} {2014}},\ \bibinfo
  {pages} {P06009} (\bibinfo {year} {2014})}\BibitemShut {NoStop}%
\bibitem [{\citenamefont {Stoudenmire}\ \emph {et~al.}(2014)\citenamefont
  {Stoudenmire}, \citenamefont {Gustainis}, \citenamefont {Johal},
  \citenamefont {Wessel},\ and\ \citenamefont {Melko}}]{Stoudenmire-2014}%
  \BibitemOpen
  \bibfield  {author} {\bibinfo {author} {\bibfnamefont {E.~M.}\ \bibnamefont
  {Stoudenmire}}, \bibinfo {author} {\bibfnamefont {P.}~\bibnamefont
  {Gustainis}}, \bibinfo {author} {\bibfnamefont {R.}~\bibnamefont {Johal}},
  \bibinfo {author} {\bibfnamefont {S.}~\bibnamefont {Wessel}}, \ and\ \bibinfo
  {author} {\bibfnamefont {R.~G.}\ \bibnamefont {Melko}},\ }\href@noop {}
  {\bibfield  {journal} {\bibinfo  {journal} {Phys. Rev. B}\ }\textbf {\bibinfo
  {volume} {90}},\ \bibinfo {pages} {235106} (\bibinfo {year}
  {2014})}\BibitemShut {NoStop}%
\bibitem [{\citenamefont {Inglis}\ and\ \citenamefont
  {Melko}(2013)}]{Inglis2013}%
  \BibitemOpen
  \bibfield  {author} {\bibinfo {author} {\bibfnamefont {S.}~\bibnamefont
  {Inglis}}\ and\ \bibinfo {author} {\bibfnamefont {R.~G.}\ \bibnamefont
  {Melko}},\ }\href@noop {} {\bibfield  {journal} {\bibinfo  {journal} {New
  Journal of Physics}\ }\textbf {\bibinfo {volume} {15}},\ \bibinfo {pages}
  {073048} (\bibinfo {year} {2013})}\BibitemShut {NoStop}%
\bibitem [{\citenamefont {Ding}\ \emph {et~al.}(2008)\citenamefont {Ding},
  \citenamefont {Bonesteel},\ and\ \citenamefont {Yang}}]{Ding2008}%
  \BibitemOpen
  \bibfield  {author} {\bibinfo {author} {\bibfnamefont {W.}~\bibnamefont
  {Ding}}, \bibinfo {author} {\bibfnamefont {N.~E.}\ \bibnamefont {Bonesteel}},
  \ and\ \bibinfo {author} {\bibfnamefont {K.}~\bibnamefont {Yang}},\
  }\href@noop {} {\bibfield  {journal} {\bibinfo  {journal} {Physical Review
  A}\ }\textbf {\bibinfo {volume} {77}},\ \bibinfo {pages} {052109} (\bibinfo
  {year} {2008})}\BibitemShut {NoStop}%
\bibitem [{\citenamefont {Metlitski}\ and\ \citenamefont
  {Grover}(2011)}]{Metlitski-2011}%
  \BibitemOpen
  \bibfield  {author} {\bibinfo {author} {\bibfnamefont {M.~A.}\ \bibnamefont
  {Metlitski}}\ and\ \bibinfo {author} {\bibfnamefont {T.}~\bibnamefont
  {Grover}},\ }\href@noop {} {\enquote {\bibinfo {title} {{Entanglement Entropy
  in Systems with Spontaneously Broken Continuous Symmetry}},}\ } (\bibinfo
  {year} {2011}),\ \Eprint {http://arxiv.org/abs/arXiv:1112.5166}
  {arXiv:1112.5166} \BibitemShut {NoStop}%
\bibitem [{\citenamefont {Ju}\ \emph {et~al.}(2012)\citenamefont {Ju},
  \citenamefont {Kallin}, \citenamefont {Fendley}, \citenamefont {Hastings},\
  and\ \citenamefont {Melko}}]{Ju2012}%
  \BibitemOpen
  \bibfield  {author} {\bibinfo {author} {\bibfnamefont {H.}~\bibnamefont
  {Ju}}, \bibinfo {author} {\bibfnamefont {A.~B.}\ \bibnamefont {Kallin}},
  \bibinfo {author} {\bibfnamefont {P.}~\bibnamefont {Fendley}}, \bibinfo
  {author} {\bibfnamefont {M.~B.}\ \bibnamefont {Hastings}}, \ and\ \bibinfo
  {author} {\bibfnamefont {R.~G.}\ \bibnamefont {Melko}},\ }\href@noop {}
  {\bibfield  {journal} {\bibinfo  {journal} {Physical Review B}\ }\textbf
  {\bibinfo {volume} {85}},\ \bibinfo {pages} {165121} (\bibinfo {year}
  {2012})}\BibitemShut {NoStop}%
\bibitem [{\citenamefont {Hastings}\ \emph {et~al.}(2010)\citenamefont
  {Hastings}, \citenamefont {Gonz\'alez}, \citenamefont {Kallin},\ and\
  \citenamefont {Melko}}]{Hastings-2010}%
  \BibitemOpen
  \bibfield  {author} {\bibinfo {author} {\bibfnamefont {M.~B.}\ \bibnamefont
  {Hastings}}, \bibinfo {author} {\bibfnamefont {I.}~\bibnamefont
  {Gonz\'alez}}, \bibinfo {author} {\bibfnamefont {A.~B.}\ \bibnamefont
  {Kallin}}, \ and\ \bibinfo {author} {\bibfnamefont {R.~G.}\ \bibnamefont
  {Melko}},\ }\href@noop {} {\bibfield  {journal} {\bibinfo  {journal}
  {Physical Review Letters}\ }\textbf {\bibinfo {volume} {104}},\ \bibinfo
  {pages} {157201} (\bibinfo {year} {2010})}\BibitemShut {NoStop}%
\bibitem [{\citenamefont {Kogut}\ and\ \citenamefont
  {Susskind}(1975)}]{Kogut-1975}%
  \BibitemOpen
  \bibfield  {author} {\bibinfo {author} {\bibfnamefont {J.}~\bibnamefont
  {Kogut}}\ and\ \bibinfo {author} {\bibfnamefont {L.}~\bibnamefont
  {Susskind}},\ }\href@noop {} {\bibfield  {journal} {\bibinfo  {journal}
  {Physical Review D}\ }\textbf {\bibinfo {volume} {11}},\ \bibinfo {pages}
  {395} (\bibinfo {year} {1975})}\BibitemShut {NoStop}%
\bibitem [{\citenamefont {Fisher}\ and\ \citenamefont
  {Fradkin}(1985)}]{Fisher-1985}%
  \BibitemOpen
  \bibfield  {author} {\bibinfo {author} {\bibfnamefont {M.~P.~A.}\
  \bibnamefont {Fisher}}\ and\ \bibinfo {author} {\bibfnamefont
  {E.}~\bibnamefont {Fradkin}},\ }\href@noop {} {\bibfield  {journal} {\bibinfo
   {journal} {Nuclear Physics B}\ }\textbf {\bibinfo {volume} {251 [FS13]}},\
  \bibinfo {pages} {457} (\bibinfo {year} {1985})}\BibitemShut {NoStop}%
\bibitem [{\citenamefont {Semenoff}(1984)}]{Semenoff-1984}%
  \BibitemOpen
  \bibfield  {author} {\bibinfo {author} {\bibfnamefont {G.~W.}\ \bibnamefont
  {Semenoff}},\ }\href@noop {} {\bibfield  {journal} {\bibinfo  {journal}
  {Physical Review Letters}\ }\textbf {\bibinfo {volume} {53}},\ \bibinfo
  {pages} {2449} (\bibinfo {year} {1984})}\BibitemShut {NoStop}%
\bibitem [{\citenamefont {Sun}\ \emph {et~al.}(2009)\citenamefont {Sun},
  \citenamefont {Yao}, \citenamefont {Fradkin},\ and\ \citenamefont
  {Kivelson}}]{Sun2009}%
  \BibitemOpen
  \bibfield  {author} {\bibinfo {author} {\bibfnamefont {K.}~\bibnamefont
  {Sun}}, \bibinfo {author} {\bibfnamefont {H.}~\bibnamefont {Yao}}, \bibinfo
  {author} {\bibfnamefont {E.}~\bibnamefont {Fradkin}}, \ and\ \bibinfo
  {author} {\bibfnamefont {S.~A.}\ \bibnamefont {Kivelson}},\ }\href@noop {}
  {\bibfield  {journal} {\bibinfo  {journal} {Physical Review Letters}\
  }\textbf {\bibinfo {volume} {103}},\ \bibinfo {pages} {046811} (\bibinfo
  {year} {2009})}\BibitemShut {NoStop}%
\bibitem [{\citenamefont {Peschel}(2003)}]{Peschel2003}%
  \BibitemOpen
  \bibfield  {author} {\bibinfo {author} {\bibfnamefont {I.}~\bibnamefont
  {Peschel}},\ }\href@noop {} {\bibfield  {journal} {\bibinfo  {journal}
  {Journal of Physics A: Theoretical and Mathematical}\ }\textbf {\bibinfo
  {volume} {36}},\ \bibinfo {pages} {L205} (\bibinfo {year}
  {2003})}\BibitemShut {NoStop}%
\bibitem [{\citenamefont {Wolf}(2006)}]{Wolf2006}%
  \BibitemOpen
  \bibfield  {author} {\bibinfo {author} {\bibfnamefont {M.}~\bibnamefont
  {Wolf}},\ }\href@noop {} {\bibfield  {journal} {\bibinfo  {journal} {Physical
  Review Letters}\ }\textbf {\bibinfo {volume} {96}},\ \bibinfo {pages}
  {010404} (\bibinfo {year} {2006})}\BibitemShut {NoStop}%
\bibitem [{\citenamefont {Gioev}\ and\ \citenamefont
  {Klich}(2006)}]{Gioev2006}%
  \BibitemOpen
  \bibfield  {author} {\bibinfo {author} {\bibfnamefont {D.}~\bibnamefont
  {Gioev}}\ and\ \bibinfo {author} {\bibfnamefont {I.}~\bibnamefont {Klich}},\
  }\href@noop {} {\bibfield  {journal} {\bibinfo  {journal} {Physical Review
  Letters}\ }\textbf {\bibinfo {volume} {96}},\ \bibinfo {pages} {100503}
  (\bibinfo {year} {2006})}\BibitemShut {NoStop}%
\bibitem [{\citenamefont {Swingle}(2010)}]{Swingle2010}%
  \BibitemOpen
  \bibfield  {author} {\bibinfo {author} {\bibfnamefont {B.}~\bibnamefont
  {Swingle}},\ }\href@noop {} {\bibfield  {journal} {\bibinfo  {journal}
  {Physical Review Letters}\ }\textbf {\bibinfo {volume} {105}},\ \bibinfo
  {pages} {050502} (\bibinfo {year} {2010})}\BibitemShut {NoStop}%
\bibitem [{\citenamefont {Ding}\ \emph {et~al.}(2012)\citenamefont {Ding},
  \citenamefont {Seidel},\ and\ \citenamefont {Yang}}]{Ding2012}%
  \BibitemOpen
  \bibfield  {author} {\bibinfo {author} {\bibfnamefont {W.}~\bibnamefont
  {Ding}}, \bibinfo {author} {\bibfnamefont {A.}~\bibnamefont {Seidel}}, \ and\
  \bibinfo {author} {\bibfnamefont {K.}~\bibnamefont {Yang}},\ }\href@noop {}
  {\bibfield  {journal} {\bibinfo  {journal} {Physical Review X}\ }\textbf
  {\bibinfo {volume} {2}},\ \bibinfo {pages} {011012} (\bibinfo {year}
  {2012})}\BibitemShut {NoStop}%
\bibitem [{\citenamefont {Calabrese}\ \emph
  {et~al.}(2012{\natexlab{b}})\citenamefont {Calabrese}, \citenamefont
  {Mintchev},\ and\ \citenamefont {Vicari}}]{Calabrese-2012}%
  \BibitemOpen
  \bibfield  {author} {\bibinfo {author} {\bibfnamefont {P.}~\bibnamefont
  {Calabrese}}, \bibinfo {author} {\bibfnamefont {M.}~\bibnamefont {Mintchev}},
  \ and\ \bibinfo {author} {\bibfnamefont {E.}~\bibnamefont {Vicari}},\
  }\href@noop {} {\bibfield  {journal} {\bibinfo  {journal} {EPL (Europhysics
  Letters)}\ }\textbf {\bibinfo {volume} {98}},\ \bibinfo {pages} {20003}
  (\bibinfo {year} {2012}{\natexlab{b}})}\BibitemShut {NoStop}%
\bibitem [{\citenamefont {Leschke}\ \emph {et~al.}(2014)\citenamefont
  {Leschke}, \citenamefont {Sobolev},\ and\ \citenamefont
  {Spitzer}}]{Leschke-2014}%
  \BibitemOpen
  \bibfield  {author} {\bibinfo {author} {\bibfnamefont {H.}~\bibnamefont
  {Leschke}}, \bibinfo {author} {\bibfnamefont {A.~V.}\ \bibnamefont
  {Sobolev}}, \ and\ \bibinfo {author} {\bibfnamefont {W.}~\bibnamefont
  {Spitzer}},\ }\href@noop {} {\bibfield  {journal} {\bibinfo  {journal}
  {Physical Review Letters}\ }\textbf {\bibinfo {volume} {112}},\ \bibinfo
  {pages} {160403} (\bibinfo {year} {2014})}\BibitemShut {NoStop}%
\bibitem [{\citenamefont {McMinis}\ and\ \citenamefont
  {Tubman}(2013)}]{McMinis2013}%
  \BibitemOpen
  \bibfield  {author} {\bibinfo {author} {\bibfnamefont {J.}~\bibnamefont
  {McMinis}}\ and\ \bibinfo {author} {\bibfnamefont {N.}~\bibnamefont
  {Tubman}},\ }\href@noop {} {\bibfield  {journal} {\bibinfo  {journal}
  {Physical Review B}\ }\textbf {\bibinfo {volume} {87}},\ \bibinfo {pages}
  {081108(R)} (\bibinfo {year} {2013})}\BibitemShut {NoStop}%
\bibitem [{\citenamefont {Witten}(1998)}]{Witten-1998}%
  \BibitemOpen
  \bibfield  {author} {\bibinfo {author} {\bibfnamefont {E.}~\bibnamefont
  {Witten}},\ }\href@noop {} {\bibfield  {journal} {\bibinfo  {journal}
  {Advances in Theoretical and Mathematical Physics}\ }\textbf {\bibinfo
  {volume} {2}},\ \bibinfo {pages} {505} (\bibinfo {year} {1998})}\BibitemShut
  {NoStop}%
\bibitem [{\citenamefont {Horowitz}\ and\ \citenamefont
  {Myers}(1998)}]{Horowitz-1998}%
  \BibitemOpen
  \bibfield  {author} {\bibinfo {author} {\bibfnamefont {G.~T.}\ \bibnamefont
  {Horowitz}}\ and\ \bibinfo {author} {\bibfnamefont {R.~C.}\ \bibnamefont
  {Myers}},\ }\href@noop {} {\bibfield  {journal} {\bibinfo  {journal}
  {Physical Review D}\ }\textbf {\bibinfo {volume} {59}},\ \bibinfo {pages}
  {026005} (\bibinfo {year} {1998})}\BibitemShut {NoStop}%
\bibitem [{\citenamefont {Lewkowycz}\ and\ \citenamefont
  {Maldacena}(2013)}]{Lewkowycz:2013nqa}%
  \BibitemOpen
  \bibfield  {author} {\bibinfo {author} {\bibfnamefont {A.}~\bibnamefont
  {Lewkowycz}}\ and\ \bibinfo {author} {\bibfnamefont {J.}~\bibnamefont
  {Maldacena}},\ }\href@noop {} {\bibfield  {journal} {\bibinfo  {journal}
  {Journal of High Energy Physics}\ }\textbf {\bibinfo {volume} {2013}},\
  \bibinfo {pages} {090} (\bibinfo {year} {2013})}\BibitemShut {NoStop}%
\bibitem [{\citenamefont {Kachru}\ \emph {et~al.}(2008)\citenamefont {Kachru},
  \citenamefont {Liu},\ and\ \citenamefont {Mulligan}}]{Kachru:2008yh}%
  \BibitemOpen
  \bibfield  {author} {\bibinfo {author} {\bibfnamefont {S.}~\bibnamefont
  {Kachru}}, \bibinfo {author} {\bibfnamefont {X.}~\bibnamefont {Liu}}, \ and\
  \bibinfo {author} {\bibfnamefont {M.}~\bibnamefont {Mulligan}},\ }\href@noop
  {} {\bibfield  {journal} {\bibinfo  {journal} {Physical Review D}\ }\textbf
  {\bibinfo {volume} {78}},\ \bibinfo {pages} {106005} (\bibinfo {year}
  {2008})}\BibitemShut {NoStop}%
\bibitem [{\citenamefont {Nishioka}\ and\ \citenamefont
  {Takayanagi}(2007)}]{Nishioka:2006gr}%
  \BibitemOpen
  \bibfield  {author} {\bibinfo {author} {\bibfnamefont {T.}~\bibnamefont
  {Nishioka}}\ and\ \bibinfo {author} {\bibfnamefont {T.}~\bibnamefont
  {Takayanagi}},\ }\href@noop {} {\bibfield  {journal} {\bibinfo  {journal}
  {Journal of High Energy Physics}\ }\textbf {\bibinfo {volume} {2007}},\
  \bibinfo {pages} {090} (\bibinfo {year} {2007})}\BibitemShut {NoStop}%
\bibitem [{\citenamefont {St{\'e}phan}\ \emph
  {et~al.}(2013{\natexlab{b}})\citenamefont {St{\'e}phan}, \citenamefont {Ju},
  \citenamefont {Fendley},\ and\ \citenamefont {Melko}}]{Stephan2013}%
  \BibitemOpen
  \bibfield  {author} {\bibinfo {author} {\bibfnamefont {J.-M.}\ \bibnamefont
  {St{\'e}phan}}, \bibinfo {author} {\bibfnamefont {H.}~\bibnamefont {Ju}},
  \bibinfo {author} {\bibfnamefont {P.}~\bibnamefont {Fendley}}, \ and\
  \bibinfo {author} {\bibfnamefont {R.~G.}\ \bibnamefont {Melko}},\ }\href@noop
  {} {\bibfield  {journal} {\bibinfo  {journal} {New Journal of Physics}\
  }\textbf {\bibinfo {volume} {15}},\ \bibinfo {pages} {015004} (\bibinfo
  {year} {2013}{\natexlab{b}})}\BibitemShut {NoStop}%
\bibitem [{\citenamefont {Ryu}\ and\ \citenamefont
  {Takayanagi}(2006{\natexlab{b}})}]{Ryu:2006ef}%
  \BibitemOpen
  \bibfield  {author} {\bibinfo {author} {\bibfnamefont {S.}~\bibnamefont
  {Ryu}}\ and\ \bibinfo {author} {\bibfnamefont {T.}~\bibnamefont
  {Takayanagi}},\ }\href@noop {} {\bibfield  {journal} {\bibinfo  {journal}
  {Journal of High Energy Physics}\ }\textbf {\bibinfo {volume} {2006}},\
  \bibinfo {pages} {045} (\bibinfo {year} {2006}{\natexlab{b}})}\BibitemShut
  {NoStop}%
\bibitem [{\citenamefont {Myers}\ and\ \citenamefont
  {Singh}(2012)}]{Myers:2012ed}%
  \BibitemOpen
  \bibfield  {author} {\bibinfo {author} {\bibfnamefont {R.~C.}\ \bibnamefont
  {Myers}}\ and\ \bibinfo {author} {\bibfnamefont {A.}~\bibnamefont {Singh}},\
  }\href@noop {} {\bibfield  {journal} {\bibinfo  {journal} {Journal of High
  Energy Physics}\ }\textbf {\bibinfo {volume} {2012}},\ \bibinfo {pages} {122}
  (\bibinfo {year} {2012})}\BibitemShut {NoStop}%
\bibitem [{\citenamefont {Potter}(2014)}]{Potter2014}%
  \BibitemOpen
  \bibfield  {author} {\bibinfo {author} {\bibfnamefont {A.}~\bibnamefont
  {Potter}},\ }\href@noop {} {\enquote {\bibinfo {title} {Boundary-law scaling
  of entanglement entropy in diffusive metals},}\ } (\bibinfo {year} {2014}),\
  \Eprint {http://arxiv.org/abs/arXiv: 1408.1094} {arXiv: 1408.1094}
  \BibitemShut {NoStop}%
\end{thebibliography}
\end{document}